\documentclass[lettersize,journal]{IEEEtran}
\usepackage{amsmath,amsfonts}
\usepackage{array}
\usepackage[caption=false,font=normalsize,labelfont=sf,textfont=sf]{subfig}
\usepackage{textcomp}
\usepackage{stfloats}
\usepackage{url}
\usepackage{verbatim}
\usepackage{graphicx}
\def\BibTeX{{\rm B\kern-.05em{\sc i\kern-.025em b}\kern-.08em
    T\kern-.1667em\lower.7ex\hbox{E}\kern-.125emX}}
\usepackage{balance}

\usepackage{cite}

\usepackage{amsmath}
\usepackage{xcolor}
\usepackage[utf8]{inputenc}
\usepackage{hyperref}

\usepackage{booktabs}
\usepackage{lipsum}
\usepackage{multirow}
\usepackage{threeparttable}
\usepackage{xcolor,colortbl}

\usepackage{graphicx}%
\usepackage{multirow}%
\usepackage{amsmath,amssymb,amsfonts}%
\usepackage{amsthm}%
\usepackage{mathrsfs}%

\usepackage{xcolor}%
\usepackage{textcomp}%
\usepackage{manyfoot}%
\usepackage{listings}%

\theoremstyle{thmstyleone}%
\newtheorem{theorem}{Theorem}

\theoremstyle{thmstyletwo}%
\newtheorem{remark}{Remark}%

\theoremstyle{thmstylethree}%
\newtheorem{definition}{Definition}%

\usepackage{caption}

\usepackage[shortlabels]{enumitem}

\usepackage[T1]{fontenc}
\usepackage{tabularx}  
\usepackage{arydshln}  
\usepackage{cryptocode}
\usepackage{xspace}
\usepackage{bm}
\usepackage{booktabs}
\usepackage{enumitem}
\usepackage{pgfplots}
\usepackage{pgfplotstable}
\usepackage{csvsimple}
\usepackage{color}
\PassOptionsToPackage{hyphens}{url}
\usepackage{hyperref}
\usepackage{cleveref}

\definecolor{darkgreen}{rgb}{0,0.4,0}

\usepackage{graphicx}%
\usepackage{multirow}%
\usepackage{amsmath,amssymb,amsfonts}%
\usepackage{amsthm}%
\usepackage{mathrsfs}%

\usepackage{xcolor}%
\usepackage{textcomp}%
\usepackage{manyfoot}%
\usepackage{algorithm}%
\usepackage{algorithmicx}%
\usepackage{algpseudocode}%
\usepackage{listings}%

\usepackage{caption}

\usepackage[shortlabels]{enumitem}

\usepackage[T1]{fontenc}
\usepackage{tabularx}  
\usepackage{arydshln}  
\usepackage{cryptocode}
\usepackage{xspace}
\usepackage{bm}
\usepackage{booktabs}
\usepackage{enumitem}
\usepackage{pgfplots}
\usepackage{pgfplotstable}
\usepackage{csvsimple}
\usepackage{color}
\PassOptionsToPackage{hyphens}{url}
\usepackage{hyperref}
\usepackage{cleveref}

\definecolor{mustardyellow}{RGB}{229, 190, 1}

\definecolor{forestgreen}{rgb}{0.0, 0.5, 0.0}

\newcommand{\updated}[1]{#1}

\usepackage{longtable}
\usepackage{booktabs}

\definecolor{clrQKD}{rgb}{1.0, 0.5, 0.0}
\definecolor{clrKC}{rgb}{0.0, 0.51, 0.53}
\definecolor{clrKPQ}{rgb}{0.28, 0.06, 0.32}
\newtheorem*{note}{Note}
\newtheorem{corr}{Corollary}
\newcommand{\VM}{\mathcmd{\text{VM}}}
\usetikzlibrary{automata}

\usetikzlibrary{matrix,positioning,chains,shapes.geometric, arrows, fit}
\tikzstyle{QKDmodule} = [rectangle, minimum width=1cm, minimum height=0.5cm, text centered, font=\scriptsize, draw=black]
\tikzstyle{KM} = [rectangle, minimum width=2cm, minimum height=0.5cm, text centered, font=\scriptsize, draw=black]
\tikzstyle{QKDN} = [rectangle, minimum width=10cm, minimum height=0.5cm, text centered, font=\scriptsize, draw=black]
\tikzstyle{arrowtwo} = [<->,>=stealth]
\tikzstyle{arrow} = [->,>=stealth]
\tikzstyle{node} = [rectangle, minimum width=2.2cm, minimum height=0.5cm, text centered, font=\scriptsize, draw=black!60, dashed]
\tikzstyle{app} = [rectangle, minimum width=2cm, minimum height=1cm, text centered, font=\scriptsize, draw=black]
\tikzstyle{desc} = [text centered, font=\scriptsize]
\tikzstyle{layer} = [rectangle ,minimum width=2cm, minimum height=0.5cm, align=right, font=\scriptsize]

\floatstyle{plain}
\newfloat{experiment}{ht}{exps}
\floatname{experiment}{Experiment}

\crefname{experiment}{experiment}{experiments}
\Crefname{experiment}{Experiment}{Experiments}

\newcommand{\INDCPA}{\ensuremath{\mathsf{IND}\text{-}\allowbreak\mathsf{CPA}}}

\newcommand{\EUFCMA}{\ensuremath{\mathsf{EUF}\text{-}\allowbreak\mathsf{CMA}}}

\newcommand{\AdvEUFCMA}[1]{\ensuremath{\mathsf{Adv}^{\mathsf{euf\text{-}cma}}_{#1}}}
\newcommand{\adv}{\advA}
\newcommand{\Adv}{\mathsf{Adv}}

\newcommand{\mathcmd}[1]{\ensuremath{#1}\xspace} 
\newcommand{\N}{\mathcmd{\mathbb{N}}}
\newcommand{\advA}{\ensuremath{\mathcal{A}}\xspace}

\newcommand{\pk}{\mathcmd{\mathsf{pk}}} 
\newcommand{\sk}{\mathcmd{\mathsf{sk}}} 
\newcommand{\setS}{\mathcmd{\mathcal{S}}} 
\newcommand{\msg}{\mathcmd{m}} 
\newcommand{\Msg}{\mathcmd{\mathcal{M}}} 

\newcommand{\key}{\mathcmd{K}} 
\newcommand{\ctxt}{\mathcmd{c}} 
\newcommand{\ExpEUFCMA}[1]{\ensuremath{\mathsf{Exp}^{\mathsf{euf\text{-}cma}}_{#1}}}
\newcommand{\Exp}{\ensuremath{\mathsf{Exp}}}
\newcommand{\ExpINDCPA}[1]{\ensuremath{\mathsf{Exp}^{\mathsf{ind\text{-}cpa}}_{#1}}}
\newcommand{\AdvINDCPA}[1]{\ensuremath{\mathsf{Adv}^{\mathsf{ind\text{-}cpa}}_{#1}}}
\newcommand{\prfeval}{\ensuremath{\mathcal{F}}}

\renewcommand{\secpar}{\kappa}
\newcommand{\kk}{\ensuremath{k}}
\newcommand{\kgen}{{\mathsf{KGen}}}
\newcommand{\GetKey}{{\mathsf{GetKey}}}
\newcommand{\ssign}{\mathsf{Sign}}
\newcommand{\sverify}{\mathsf{Ver}}
\newcommand{\DSIG}{{\ensuremath{\mathsf{DSS}}}}
\newcommand{\psk}{\mathcmd{\mathsf{psk}}}
\newcommand{\Key}{\mathcmd{\mathcal{K}}}
\newcommand{\KEM}{\mathcmd{\mathsf{KEM}}} 

\newcommand{\KEMINDCPA}{IND-CPA\xspace}

\newcommand{\ExpKEMINDCPA}[1]{\ensuremath{\mathsf{Exp}^{\mathsf{ind\text{-}cpa}}_{#1}}}

\newcommand{\Encaps}{\mathcmd{\mathsf{Enc}}} 
\newcommand{\Decaps}{\mathcmd{\mathsf{Dec}}} 

\newcommand{\MAC}{\mathcmd{\mathsf{MAC}}}
\newcommand{\msign}{\mathsf{Auth}}
\newcommand{\mverify}{\mathsf{Ver}}

\newcommand{\SigMuckle}{\mathcmd{\text{Muckle}+}}

\newcommand{\VMuckle}{\mathcmd{\text{VMuckle}}}

\newcommand{\lbl}{\mathcmd{\ell}}
\newcommand{\SecState}{\mathcmd{\mathsf{SecState}}}

\newcommand{\clean}{\mathcmd{\mathsf{clean}}}
\newcommand{\Corrupt}{\mathcmd{\mathsf{Corrupt}}}

\newcommand{\CorruptSK}{\mathcmd{\mathsf{CorruptSK}}}
\newcommand{\CorruptQK}{\mathcmd{\mathsf{CorruptQK}}}
\newcommand{\CorruptCK}{\mathcmd{\mathsf{CorruptCK}}}
\newcommand{\Compromise}{\mathcmd{\mathsf{Compromise}}}
\newcommand{\CompromiseSS}{\mathcmd{\mathsf{CompromiseSS}}}
\newcommand{\CompromiseSK}{\mathcmd{\mathsf{CompromiseSK}}}
\newcommand{\CompromiseQK}{\mathcmd{\mathsf{CompromiseQK}}}
\newcommand{\CompromiseCK}{\mathcmd{\mathsf{CompromiseCK}}}
\newcommand{\Test}{\mathcmd{\mathsf{Test}}}
\newcommand{\Create}{\mathcmd{\mathsf{Create}}}
\newcommand{\Reveal}{\mathcmd{\mathsf{Reveal}}}
\newcommand{\Send}{\mathcmd{\mathsf{Send}}}
\newcommand{\act}{\mathcmd{\mathsf{active}}}
\newcommand{\acc}{\mathcmd{\mathsf{accept}}}
\newcommand{\rej}{\mathcmd{\mathsf{reject}}}

\begin{document}

\title{Versatile Quantum-Safe Hybrid Key Exchange and Its Application to MACsec}

\author{
    Jaime~S.~Buruaga\IEEEauthorrefmark{1},
    Augustine~Bugler\IEEEauthorrefmark{2}, 
    Juan~P.~Brito\IEEEauthorrefmark{3}, 
    Vicente~Martin\IEEEauthorrefmark{3},
    Christoph~Striecks\IEEEauthorrefmark{4}\\
    \IEEEauthorblockA{\IEEEauthorrefmark{1}Center for Computational Simulation, Universidad Polit\'ecnica de Madrid, Madrid, Spain}\\
    \IEEEauthorblockA{\IEEEauthorrefmark{2}University of Vienna, Vienna, Austria}\\
    \IEEEauthorblockA{\IEEEauthorrefmark{3}Center for Computational Simulation and DLSIIS, ETSI Inform\'aticos, Universidad Polit\'ecnica de Madrid, Madrid, Spain}\\
    \IEEEauthorblockA{\IEEEauthorrefmark{4}AIT Austrian Institute of Technology, Vienna, Austria}\\
    \IEEEcompsocitemizethanks{\IEEEcompsocthanksitem Jaime~S.~Buruaga is the corresponding author: j.saezdeburuaga@upm.es. \IEEEcompsocthanksitem During the execution of this research, Augustine Bugler was affiliated with AIT.}
}

\maketitle

\begin{abstract}
Advancements in quantum computing pose a significant threat to most of the cryptography currently deployed. Fortunately, cryptographic building blocks to mitigate the threat are already available; mostly based on post-quantum and quantum cryptography, but also on symmetric cryptography techniques. Notably, quantum-safe building blocks must be deployed as soon as possible due to the ``harvest-now decrypt-later'' attack scenario, which is already challenging our sensitive and encrypted data today.

Following an agile defense-in-depth approach, Hybrid Authenticated Key Exchange (HAKE) protocols have recently been gaining significant attention. Such protocols modularly combine conventional, post-quantum, and quantum cryptography to achieve confidentiality, authenticity, and integrity guarantees for network channels. Unfortunately, only a few protocols have yet been proposed (mainly Muckle and \SigMuckle) with different flexibility guarantees.

Looking at available standards in the network domain (especially at the Media Access Control Security (MACsec) standard), we believe that HAKE protocols could already bring strong security benefits to MACsec today. MACsec is a standard designed to secure communication at the data link layer in Ethernet networks by providing security for all traffic between adjacent entities. In addition, MACsec establishes secure channels within a Local Area Network (LAN), ensuring that data remain protected from eavesdropping, tampering, and unauthorized access, while operating transparently to higher layer protocols. Currently, MACsec does not offer enough protection in the event of cryptographically relevant quantum computers.

In this work, we tackle the challenge and propose a new versatile HAKE protocol, dubbed \VMuckle, which is sufficiently flexible for the use in MACsec to provide LAN participants with hybrid key material ensuring secure communication.
\end{abstract}

\begin{IEEEkeywords}
Quantum Key Distribution, Post-Quantum Cryptography, Transport Security Layer, Hybrid Authenticated Key Exchange, Software-Defined Networking, Crypto-agility, Performance evaluation
\end{IEEEkeywords}

\section{Introduction}\label{sec:Introduction}
Quantum computing, particularly through the implementation of Shor's algorithm \cite{Shor}, poses a significant threat to contemporary cryptography systems, such as RSA, Diffie-Hellman Key Exchange (DHKE), and Elliptic-curve Cryptography (ECC). This is because Shor's algorithm efficiently solves the problems of integer factorization and discrete logarithm, effectively breaking the mathematical foundations underpinning the security of these cryptographic schemes in the classical computing paradigm. To address this emerging threat, two principal solutions have been proposed: Post-Quantum Cryptography (PQC) and Quantum Key Distribution (QKD).

Since 2016, NIST has been leading an initiative to identify new public key cryptographic algorithms that are resistant to quantum computing attacks, with the goal of replacing the ones currently in use. This process culminated in 2022 \cite{NIST_8413} and resulted in the standardization in 2024 of three algorithms: a Key Encapsulation Mechanism (ML-KEM \cite{FIPS203}) and two digital signature schemes (ML-DSA \cite{FIPS204} and SLH-DSA \cite{FIPS205}). While these algorithms demonstrate resistance to Shor's algorithm, their resilience against future quantum attacks remains unproven. \updated{By using these new algorithms, the quantum-safe PKIs can be constructed by replacing the old public key algorithms. This also has the advantage that all management procedures for key revocation and certificate distribution remain the same.}

QKD leverages quantum mechanics to establish shared keys between two parties \cite{QKD}. Unlike PQC or classical cryptography, which rely on the computational difficulty of mathematical problems, QKD's security is rooted in the fundamental principles of quantum physics. This inherent property ensures its immunity to computational attacks, eliminating the need for assumptions about the computational power of potential adversaries. Consequently, QKD provides a more predictable long-term security profile compared to computationally-based cryptographic methods, whose security may degrade with advances in computing, cryptanalysis, mathematical theory or even the emergence of new computational paradigms, as has been the case with quantum computing.

Hybridizing PQC and QKD is crucial for creating quantum-safe solutions that ensure crypto-agility, a fundamental requirement for future-proof security frameworks. PQC provides cryptographic algorithms that are resistant to quantum attacks, but its security is still based on mathematical assumptions that could be compromised by advances in quantum algorithms or unforeseen vulnerabilities \cite{PQC}. In contrast, QKD guarantees Information-Theoretic Security (ITS) \cite{Scarani}, offering unconditional security. However, QKD is constrained by practical limitations, such as transmission distance and implementation challenges. By combining PQC with QKD, we can leverage the scalability of PQC and the absolute security of QKD, creating systems that are resilient to both classical and quantum threats. This hybrid approach not only enhances security, but also facilitates crypto-agility by enabling the dynamic integration of cryptographic techniques as technology evolves, ensuring long-term confidentiality, integrity, and authentication. 

Authenticated Key Exchange (AKE) protocols are the main building blocks for secure communication channels to guarantee confidentiality, integrity, and authenticity between two entities in a network. Hybrid AKE (HAKE) protocols extend this concept, enabling the distribution of PQC, QKD, and classical key material in an authenticated manner. Although traditionally deployed at higher layers of the network stack, this work adapts HAKE protocols for efficient use at lower network layers.

\updated{Ensuring quantum-safe security for authentication processes is a critical task, and it should be extended to a broad range of services across the current telecommunication networks -- not only for MACsec, but also for other network protocols and technologies such as IPsec \cite{QS_MACsec}, TLS \cite{TLS_ArXiV}, or Blockchain \cite{YAF24, YSJ22}.} 

Media Access Control security (MACsec), defined in the IEEE 802.1AE standard \cite{IEEE802.1AE-2018}, provides robust point-to-point security on Ethernet links. Operating at layer 2 of the OSI model \cite{OSI} (data link layer), MACsec ensures the integrity, confidentiality, and authenticity of Ethernet frames.  MACsec offers device-to-device security by establishing a secure channel between the MACsec stations regardless of the number of intervening devices or networks, allowing it to secure data communication in Local Area Networks (LANs), Metropolitan Area Network (MANs) or Wide Area Network (WANs).

From a performance perspective, MACsec is a highly efficient protocol due to its low-latency characteristics, achieved through hardware-based processing. MACsec provides high-speed transparent encryption across individual network links without affecting higher-layer protocols. Its key benefits include flexibility in supporting both unicast and multicast traffic, seamless integration with existing network infrastructure, and scalability to accommodate a wide range of network topologies and sizes. Moreover, its transparency to network applications allows any application to operate over a MACsec-protected network without the need for modifications, thereby ensuring robust security with minimal impact on network performance and application compatibility.

Although MACsec employs AES-256 for payload encryption, which provides strong resistance to quantum attacks \cite{AES_QR}, the initial establishment of these symmetric keys depends on Pre-Shared Keys (PSKs) or the EAP-TLS protocol, as described in the authentication process defined by the IEEE 802.1X standard \cite{IEEE802.1X}. The former does not provide scalability in larger networks, while reliance on the latter introduces a vulnerability, as EAP-TLS employs classical asymmetric cryptography for key exchange, making it susceptible to quantum attacks.

In recent years, there has been an increasing trend towards integrating post-quantum authentication and key exchange mechanisms into major security protocols used in telecommunications networks. As MACsec serves as the preferred protocol for securing the data link layer in modern communication infrastructures, it is of critical importance to ensure that its operation is conducted in a quantum-resistant manner.

This work introduces a quantum-safe authenticated key agreement protocol for MACsec network nodes. This key can then be utilized by the MACsec Key Agreement (MKA) protocol as the root key in its key hierarchy to derive secure keys for the execution of MACsec sessions. To achieve this objective, a HAKE protocol has been developed. This protocol enables secure root key agreement using a combination of classical, post-quantum, and quantum techniques, ensuring crypto-agility and long-term confidentiality, mitigating vulnerabilities arising from future cryptographic advancements, and safeguarding the integrity of key exchanges.

To establish a shared authenticated key between any two entities, it is tempting to use ideas from the well-established Transport Layer Security (TLS) protocol. While there is indeed some work on hybrid TLS 1.3 \cite{Stebila_24}, no protocol with quantum-safe provable security guarantees for TLS is yet available. Hence, in this work, we will instead follow the HAKE framework, which is capable of providing a proof of security for our envisioned key exchange mechanism.

More concretely, we propose \VMuckle, a new adaptation of the \SigMuckle HAKE protocol \cite{BruRamStr23} tailored for small-- and large--scale quantum-safe environments. To achieve their desired security properties, we modify the protocol flow and proof techniques, proving its security within the HAKE framework. \updated{The result is that we have (quantum-safe) security if at least one authentication method and one key component are (quantum-safe) secure and have not been compromised. In particular, this is what allows for the versatility of authentication methods in $\VMuckle$ in comparison to previous HAKE protocols.}

Additionally, we demonstrate how to seamlessly replace 802.1X authentication with \VMuckle to establish secure, authenticated keys for MKA, without modifying the 802.1AE or 802.1X standards.
This versatile solution integrates classical, post-quantum, and quantum key distribution methods, supporting authentication through PSKs, post-quantum digital signatures, or both, to provide MKA with a secure root key.

This work ensures that MACsec remains a future-proof standard, capable of addressing both classical and quantum security threats.

\section{Background}\label{sec:Background}

\subsection{MACsec}\label{subsec:MACsec}

802.1AE does not specify procedures for key management or distribution within a MACsec-secured network, leaving the responsibility of authentication and key provisioning to network administrators. However, 802.1X provides a robust framework for authenticating devices and granting or denying access to network resources, handling device authentication through methods such as EAP-TLS or PSKs. Once device authentication is completed, the MKA protocol, also defined in 802.1X, manages key distribution and secure communication channel establishment. MKA automates the configuration and maintenance of these secure Layer 2 communication channels between authenticated devices, facilitating key distribution, synchronization, and association of cryptographic parameters required for MACsec to operate effectively.

A Connectivity Association (CA) is a security relationship that comprises a fully connected subset of the service access points at stations attached to a single LAN, MAN, or WAN, which is supported by MACsec. It serves as a broader security construct that encompasses multiple unidirectional Security Associations (SAs), thereby enabling a cohesive and secure communication environment between devices within a network domain. A CA is composed of one or more Secure Channels (SC), where each SC is identified by a unique Secure Channel Identifier (SCI). These SCs are formed by pairs of unidirectional SAs, enabling bidirectional communication. CAs can support both unicast and multicast scenarios and are secured using a Connectivity Association Key (CAK), which is a shared secret known by all CA participants identified by the Connectivity Association Key Name (CKN). The CAK represents the root of the MKA key hierarchy and is used in the derivation of both the Integrity Check Key (ICK) and the Key Encryption Key (KEK).

\begin{figure}[htbp]
\centering
    \includegraphics[width=.97\linewidth]{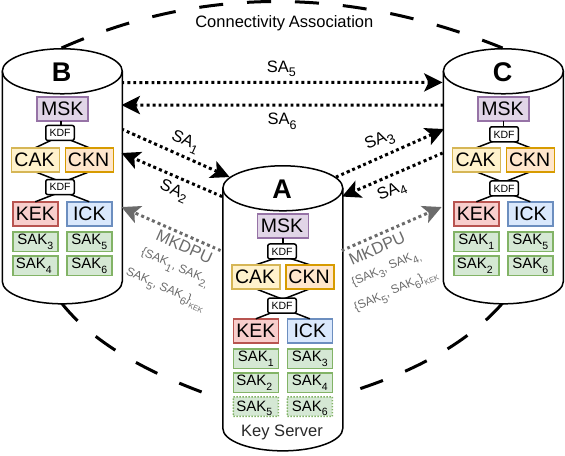}
\caption{MACsec Key Agreement protocol operation in a 3-node scenario with any-to-any Security Associations deployment (see text for further explanation).}
\label{fig:MACsec}
\end{figure}

Figure \ref{fig:MACsec} illustrates the creation and management of a CA handled by the MKA protocol. Initially, all CA participants are authenticated using 802.1X, resulting in the generation of a Master Session Key (MSK), which will be used to derive the CAK and CKN. Using these keys, any node can derive the KEK and ICK. Next, a station is elected as the key server, responsible for generating any required Secure Association Key (SAK). This selection is performed dynamically among all CA participants by generating a message that includes key server priority and a list of active nodes. The node with the highest priority is assigned as the key server, and retains this role until it is no longer present in the list of active nodes. In the depicted scenario, an SA is created for any-to-any communication, leading to the generation of six SAKs. The MKA protocol coordinates the management and distribution of these keys within the CA through MKA Data Protocol Unit (MKDPU) frames to communicate their identity, role, and availability of new SAKs to the nodes involved (i.e., SAK$_1$ and SAK$_2$ for A-B communication, SAK$_3$ and SAK$_4$ for A-C communication, and SAK$_5$ and SAK$_6$ for B-C communication). The confidentiality of these frames is ensured by the encryption with the KEK and the integrity by the ICK. At this point, all nodes are ready to initiate secure MACsec sessions with any peer in the CA. The standard allows dynamic maintenance of the CA to adapt to changes in the network topology or membership. If a new device joins or an existing device leaves the CA, the MKA protocol seamlessly manages these transitions, updating the SAs as needed.

\subsection{Main cryptographic protocols}

The main building block of secure communication between two entities is an authenticated key exchange (AKE). Recently, hybrid AKE (HAKE) has been receiving significant attention due to their ability to derive a shared authenticated key from several key-material sources (i.e., from classical and post-quantum keys, as well as from QKD).

Dowling, Brandt Hansen and Paterson \cite{DowHanPat20} proposed the first work in the area and gave an instantiation which they dubbed Muckle. For authentication, Muckle uses PSKs, but this method does not scale for large networks. However, in smaller quantum-safe networks, where PSKs can be distributed readily, Muckle may be used.

In a subsequent work, Bruckner, Ramacher and Striecks \cite{BruRamStr23} addressed the gap regarding larger-scale networks and proposed $\SigMuckle$, which utilizes post-quantum signatures for the authentication of the entities, resulting in a very efficient way to verify the key exchange. As it turns out, \SigMuckle is particularly interesting for large-scale quantum-safe networks since the scheme no longer relies on PSKs. A requirement of \SigMuckle is the deployment of a Public Key Infrastructure (PKI).

Notably, smaller networks with only a few entities may not always require a full-blown PKI, and in such cases it may suffice to utilize PSKs for the authentication. However, larger networks, such as the anticipated European Quantum Communication Infrastructure (EuroQCI) \cite{EuroQCI}, would likely benefit from also deploying a PKI due to the expected large number of communicating entities.

Unfortunately, neither of the two HAKE protocols, Muckle and \SigMuckle, is versatile in the sense that they do not allow both authentication mechanisms (i.e., post-quantum signatures or PSKs) to be chosen adaptively during the protocol run. In this work, we address this gap and propose \VMuckle to overcome these limitations. Particularly, this results in a flexible use of HAKE protocols suited for both small-- and large--scale quantum-safe networks. Our approach in this work yields a modular integration of \VMuckle into MACsec such that both quantum-safe authentication options are now available with provable security guarantees.

\subsection{Related Work}

The increasing concern regarding potential quantum attacks has intensified the emphasis on enhancing network security, prompting various initiatives to incorporate quantum-safe solutions into established network security protocols, such as MACsec. This section provides a comprehensive overview of these efforts, outlining the current state of post-quantum integration and highlighting the advancements introduced by our approach in comparison to existing methodologies.

The authors of \cite{PQC_MACsec} propose an authenticated post-quantum key establishment protocol aimed at securing long-term MACsec sessions by providing an ephemeral session key exchange mechanism capable of deriving an encryption key directly from a post-quantum public key scheme. Their approach leverages the Classic McEliece public key cryptosystem as a Key Encapsulation Mechanism (KEM), integrated within the EAP-TLS authentication framework.

The same authors present an additional study that investigates the use of QKD as an alternative source of trust for MACsec \cite{QKD_MACsec}. \updated{However, their proposal does not make use of the 802.1X standard, the MKA protocol, it only uses an extension of the SAK by mixing a QKD key with the standard Diffie-Hellman.}

\updated{Stefan-Lukas Gazdag \textit{et al.} (\cite{QS_MACsec}, Sect. 5.1) outline two approaches to achieve a quantum-secure MKA. First, for the standard hierarchical mode, they introduce EAP‑TLS‑PQ, mandating post‑quantum ciphersuites that use PQ certificates and a PQ key exchange. Second, for an ephemeral mode, they design an AKE that blends a post‑quantum exchange with classical cryptography to produce a hybrid key. Both paths rely on ciphersuites and fit cleanly into existing MACsec workflows. They use Classic McEliece, NTRU, CRYSTALS‑Kyber and SABER alongside DHKE for key exchange, and CRYSTALS‑Dilithium and Falcon for authentication. In this work, we follow the same approach in the ephemeral mode, but we extend the protection of the mechanism by using classical, post-quantum and quantum cryptography underpinning our claims with a proof of security (where latter is currently still absent for EAP‑TLS‑PQ).} 

Our work extends these approaches by introducing a quantum-safe HAKE protocol, designed to provide the MKA protocol with an authenticated root key for the key hierarchy \updated{achieved through the combination of recently approved PQC algorithms (\cite{FIPS203, FIPS204, FIPS205}) together with QKD and current classical cryptography}. This innovation enables the establishment of a hybrid quantum-resistant MACsec session, ensuring enhanced security against quantum attacks, and security fallback in case any of the proposed methods is compromised in the future.

\section{$\VMuckle$: A Versatile Key Exchange Protocol}\label{sec:vmuckle}

In this section, we provide a Hybrid Authenticated Key Exchange protocol, dubbed $\VMuckle$, that is used to substitute the 802.1X authentication in MACsec. $\VMuckle$ combines key material from classical and post-quantum mechanisms, as well as from QKD. Authentication can be achieved via Pre-Shared Keys, or a post-quantum authentication mechanism.

As building blocks, we use a pseudo-random function (PRF), a message authentication code (MAC), a digital signature scheme (DSS), key encapsulation mechanisms (KEMs), and an authenticated encryption scheme with associated data (AEAD). (These are defined in Appendix \ref{app:pre-hake}, together with further relevant cryptographic security aspects.)

In the following section, we describe the protocol flow of \VMuckle. In the initial stage, the initiator and the responder possess certificates $cert_I$ and $cert_R$, respectively, from a Public Key Infrastructure (the details of which are out of the scope of this work). The certificates contain public verification keys $\pk_I$ and $\pk_R$, and the initiator and the responder possess the corresponding private keys $\sk_I$ and $\sk_R$, respectively.

The initiator and the responder also possess a shared secret state $\SecState$ (which is initially set to an empty string). They additionally have access to the labels defined in \Cref{tab:labels}. These labels are used to ensure separation in the key derivation function (KDF). 

The initiator chooses a random bitstring $n_I$ with bit-length $\secpar$, where $\secpar$ is at least 128, and generates ephemeral public-secret key pairs $(\pk_{c}, \sk_{c})$, $(\pk_{pq}, \sk_{pq})$ via the key generation algorithms of the classical and post-quantum KEMs, respectively. (If no classical key pair $(\pk_c, \sk_c)$ is generated, then the initiator sets $\pk_c$ and $\sk_c$ to empty strings.)

\begin{table}[t]
\centering
\caption{Values for the contexts used in the \updated{\VMuckle} key schedule. The context inputs follow the choices in the TLS 1.3 handshake~\cite{JC:DFGS21}.}
\label{tab:contexts}
\begin{tabular}{c|l|c|l}
    \toprule
    Label & Context Input & Label & Context Input \\
    \midrule
    $H_\varepsilon$ & ``'' &
    $H_0$ & $H(\text{``''})$ \\
    $H_1$ & $H(\bm{m_1} \| \bm{m_2})$ &
    $H_2$ & $H(\bm{m_1} \| \ldots \| \bm{m_3})$ \\
    $H_3$ & $H(\bm{m_1} \| \ldots \| \bm{m_4})$  &
    $H_4$ & $H(\bm{m_1} \| \ldots \| \bm{m_5})$  \\
    $H_5$ & $H(\bm{m_1} \| \ldots \| \bm{m_6})$  &
    $H_6$ & $H(\bm{m_1} \| \ldots \| \bm{m_7})$  \\
    \bottomrule
\end{tabular}
\end{table}

\begin{table}[t]
\centering
\caption{Values for the labels used in the \VMuckle (taken from \SigMuckle \cite{BruRamStr23}).}
\label{tab:labels}

\resizebox{\columnwidth}{!}{
\begin{tabular}{c|l|c|l}
    \toprule
    Label & Label Input & Label & Label Input \\
    \midrule
    $\lbl_0$ & ``derive k c'' &
    $\lbl_1$ & ``derive k pq'' \\
    $\lbl_2$ & ``first ck'' &
    $\lbl_3$ & ``second ck'' \\
    $\lbl_4$ & ``third ck'' &
    $\lbl_5$ & ``fourth ck'' \\
    $\lbl_6$ & ``derived'' &
    $\lbl_7$ & ``c hs traffic'' \\
    $\lbl_8$ & ``s hs traffic'' &
    $\lbl_9$ & ``finished''\\
    $\lbl_{10}$ & ``c ap traffic'' &
    $\lbl_{11}$ & ``s ap traffic''\\
    $\lbl_{12}$ & ``secstate'' &
    $\lbl_{13}$ & ``TLS 1.3, server CertificateVerify'' \\
    $\lbl_{14}$ & ``TLS 1.3, client CertificateVerify'' & \\
    \bottomrule
\end{tabular}
}
\end{table}

The initiator then sends the public keys and the random string to the responder via a public channel.

The responder chooses a random bitstring $n_R$ with bit-length $\secpar$. The responder generates an encapsulation $(\ctxt_{pq}, ss_{pq})$ via the encapsulation algorithm of the post-quantum KEM, using the public key $\pk_{pq}$. If $\pk_c$ is available, the responder also generates an encapsulation $(\ctxt_c, ss_c)$ via the encapsulation algorithm of the classical KEM, using the public key $\pk_c$. If $\pk_c$ is an empty string, then $\ctxt_c$ and $ss_c$ are set as empty strings. The responder sends $\ctxt_{pq}$, $\ctxt_c$, and $n_R$ to the initiator via a public channel. 

The initiator now computes the decapsulation $ss_{pq}$ via the decapsulation algorithm of the post-quantum KEM using the secret key $\sk_{pq}$ with the ciphertext $\ctxt_{pq}$. If $\ctxt_c$ is non-empty, the decapsulation of $ss_c$ is also computed using the decapsulation algorithm of the classical KEM with the secret key $\sk_c$ and the ciphertext $\ctxt_c$. If $\ctxt_c$ is an empty string, then $ss_c$ is set as an empty string. 

\begin{figure*}[htbp]
\centering
\fontsize{6.6pt}{6.6pt}\selectfont
\begin{tikzpicture}[outer ysep = 1mm]
  \node(1){
    \begin{tabular}{p{4.5cm} >{\centering}p{2.5cm} >{\raggedleft}p{4.5cm}}
        \centering Initiator & & \centering Responder \tabularnewline
        \midrule
        \centering $\sk_I, \pk_R$, $cert_I$, \boxed{\psk}, \SecState, $l_0,\ldots,l_{14}$, $H_\varepsilon,H_0,\ldots,H_{6}$  & & \centering $\sk_R, \pk_I$, $cert_R$, \boxed{\psk}, \SecState, $l_0,\ldots,l_{14}$, $H_\varepsilon,H_0,\ldots,H_{6}$ \tabularnewline
        & & \tabularnewline
        
        $n_I \gets \{0,1\}^{\secpar}$ & & \tabularnewline
        $\pk_c, \sk_c \gets \KEM_c.\kgen(\secpar)$ & & \tabularnewline
        $\pk_{pq}, \sk_{pq} \gets \KEM_{pq}.\kgen(\secpar)$ & & \tabularnewline
        & $\xrightarrow{\makebox[2cm]{$\bm{m_1}\colon \pk_c, \pk_{pq}, n_I$}}$ & $n_R \gets \{0,1\}^{\secpar}$ \tabularnewline
        & & $\ctxt_c, ss_c \gets \KEM_c.\Encaps(\pk_c)$ \tabularnewline
        & & $\ctxt_{pq}, ss_{pq} \gets \KEM_{pq}.\Encaps(\pk_{pq})$ \tabularnewline
        & $\xleftarrow{\makebox[2cm]{$\bm{m_2}\colon \ctxt_c, \ctxt_{pq}, n_R$}}$ & \tabularnewline
        $ss_c \gets \KEM_c.\Decaps(\sk_c, \ctxt_c)$ & & \tabularnewline
        $ss_{pq} \gets \KEM_{pq}.\Decaps(\sk_{pq}, \ctxt_{pq})$ & & \tabularnewline
        & & \tabularnewline
        
        \multicolumn{3}{c}{$k_c \gets \prfeval(ss_c, \lbl_0 \| H_1)$}  \tabularnewline
        \multicolumn{3}{c}{$k_{pq} \gets \prfeval(ss_{pq}, \lbl_1 \| H_1)$} \tabularnewline
        & & \tabularnewline
        \multicolumn{3}{c}{$k_0 \gets \prfeval(k_{pq}, \lbl_2 \| H_1)$}  \tabularnewline
        \multicolumn{3}{c}{$k_1 \gets \prfeval(k_c, \lbl_3 \| k_0)$} \tabularnewline
        & & \tabularnewline
        $k_q\gets\GetKey_{qkd}(\secpar)$ & & $k_q\gets\GetKey_{qkd}(\secpar)$\tabularnewline
        \multicolumn{3}{c}{$k_2 \gets \prfeval(k_{q}, \lbl_4 \| k_1)$} \tabularnewline
        \multicolumn{3}{c}{$k_3 \gets \prfeval(\SecState, \lbl_5 \| k_2)$}
´    \end{tabular}
  };
  \node(2)[below=0cm of 1]{
    \begin{tabular}{p{4.5cm} >{\centering}p{2.5cm} >{\raggedleft}p{4.5cm}}
        \multicolumn{3}{c}{$CHTS \gets \prfeval(k_3, \lbl_7 \| H_1)$} \tabularnewline
        \multicolumn{3}{c}{$SHTS \gets \prfeval(k_3, \lbl_8 \| H_1)$} \tabularnewline
        \multicolumn{3}{c}{$dHS \gets \prfeval(k_3, \lbl_6 \| H_0)$} \tabularnewline
        \multicolumn{3}{c}{$tk_{chs} \gets \prfeval(CHTS)$} \tabularnewline
        \multicolumn{3}{c}{$tk_{shs} \gets \prfeval(SHTS)$} \tabularnewline
        & & \tabularnewline
        \multicolumn{3}{c}{$fk_C \gets \prfeval(CHTS, \lbl_9 \| H_\varepsilon)$} \tabularnewline
        \multicolumn{3}{c}{$fk_S \gets \prfeval(SHTS, \lbl_9 \| H_\varepsilon)$} \tabularnewline
        & & \tabularnewline

        Verify $cert_R$ & $\xleftarrow{\makebox[2cm]{$\bm{m_3}\colon \{ cert_R \}_{tk_{shs}}$}}$ & \tabularnewline
        & & $\sigma_R \gets \DSIG.\ssign(\sk_R, \lbl_{13} \| H_2)$ \tabularnewline
        $\DSIG.\sverify(\pk_R, \lbl_{13} \| H_2, \sigma_R) \stackrel{?}{=} 1$ & $\xleftarrow{\makebox[2cm]{$\bm{m_4}\colon \{\sigma_R\}_{tk_{shs}}$}}$ & \tabularnewline
        & & $\tau_R \gets \MAC.\msign(\boxed{\prfeval(\psk,fk_S)}, H_3)$ \tabularnewline
        $\MAC.\mverify(\boxed{\prfeval(\psk,fk_S)}, H_3, \tau_R) \stackrel{?}{=} 1$ & $\xleftarrow{\makebox[2cm]{$\bm{m_5}\colon \{\tau_R\}_{tk_{shs}}$}}$ & \tabularnewline
        & & \tabularnewline

        \multicolumn{3}{c}{$MS \gets \prfeval(dHS, 0)$} \tabularnewline
        \multicolumn{3}{c}{$CATS \gets \prfeval(MS, \lbl_{10} \| H_4)$} \tabularnewline
        \multicolumn{3}{c}{$SATS \gets \prfeval(MS, \lbl_{11} \| H_4)$} \tabularnewline
        \multicolumn{3}{c}{$\SecState \gets \prfeval(MS, \lbl_{12} \| H_4)$} \tabularnewline
        & & \tabularnewline

        & $\xrightarrow{\makebox[2cm]{$\bm{m_6}\colon \{cert_I\}_{tk_{chs}}$}}$ & Verify $cert_I$ \tabularnewline
        $\sigma_I \gets \DSIG.\ssign(\sk_I, \lbl_{14} \| H_5)$ & & \tabularnewline
        & $\xrightarrow{\makebox[2cm]{$\bm{m_7}\colon \{\sigma_I\}_{tk_{chs}}$}}$  & $\DSIG.\sverify(\pk_I, \lbl_{14} \| H_5, \sigma_I) \stackrel{?}{=} 1$  \tabularnewline
        $\tau_I \gets \MAC.\msign(\boxed{\prfeval(\psk,fk_C)}, H_6)$  &  \tabularnewline
        & $\xrightarrow{\makebox[2cm]{$\bm{m_8}\colon \{\tau_I\}_{tk_{chs}}$}} $ & $\MAC.\mverify(\boxed{\prfeval(\psk,fk_C)}, H_6, \tau_I) \stackrel{?}{=} 1$
    \end{tabular}
  }; 

  \draw[dotted](1.south west) -- (1.south east) node[midway, fill=white] {\textit{end of ephemeral key exchange phase}};
  \draw[dotted](2.south west) -- (2.south east) node[midway, fill=white] {\textit{end of authentication and key confirmation phase}};
    
\end{tikzpicture}
{\captionsetup{font=small}
\caption{\updated{One stage of the \VMuckle protocol with a classical KEM $\KEM_c$, a post-quantum KEM $\KEM_{pq}$, a MAC $\MAC$ and a PRF $\prfeval$, where $k_q$ is a symmetric QKD key (provided out-of-band via the function $\GetKey_{qkd}$), and $\psk$ is a pre-shared key. $cert_I$ and $cert_R$ are certificates (provided out-of-band) for the public keys $\pk_I$ and $\pk_R$, respectively. $CATS$ and $SATS$ are the client and server application traffic secrets, respectively. Messages $\bm{m_i} : \{ \msg_{i,1}, \ldots \}_k$ denote that $\msg_{i,1}, \ldots$ are encrypted with an authenticated encryption scheme using the secret key $k$. The various contexts and labels are provided in \Cref{tab:contexts,tab:labels}. In the first run, \SecState is initialized as an empty string $\bot$. Changes to \SigMuckle are denoted $\boxed{\text{boxed}}$.}}}
\label{fig:sig_muckle_stage}
\end{figure*}
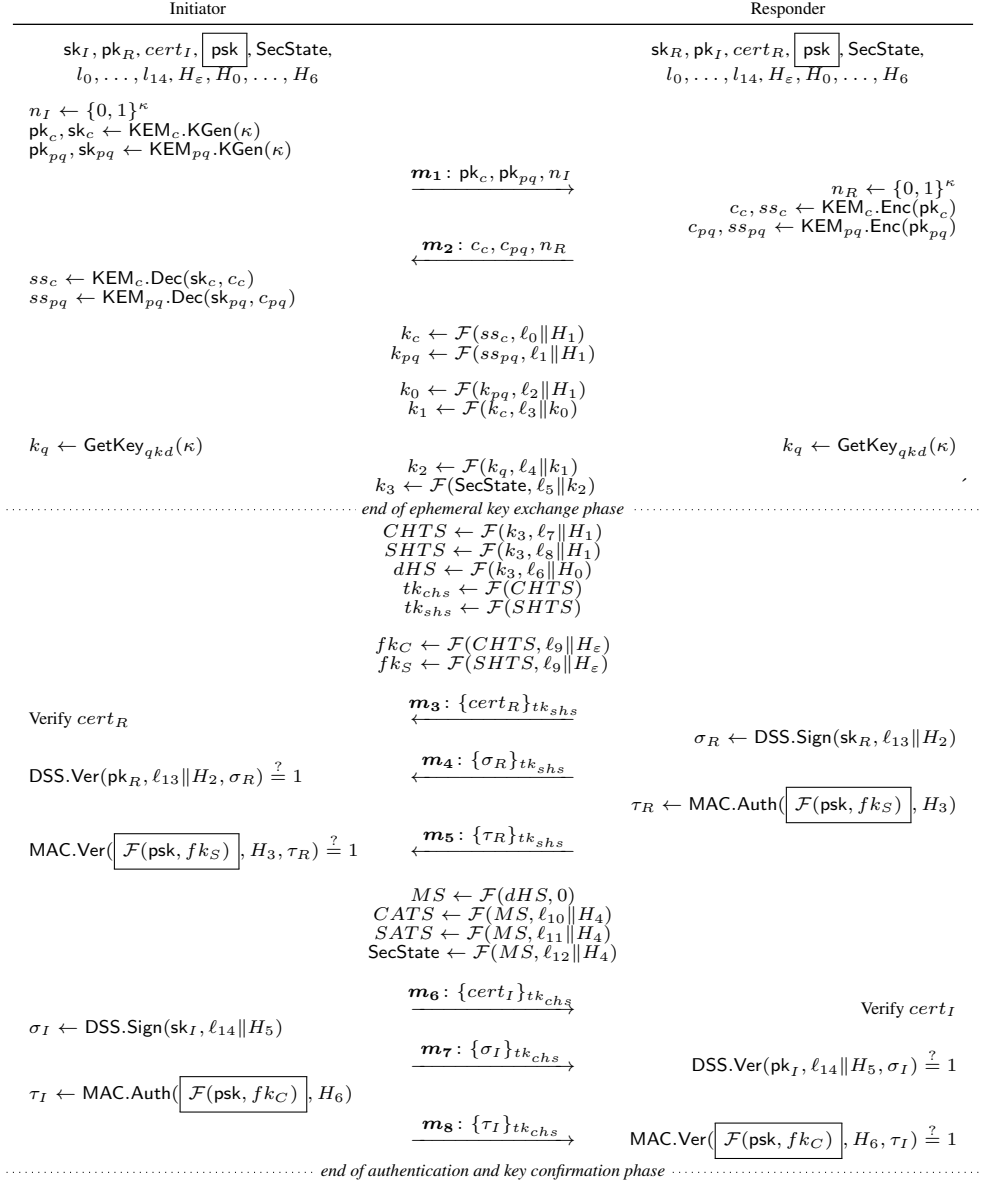

Next, the initiator and responder invoke a series of KDF calls as given in \Cref{fig:sig_muckle_stage}, where $k_q$ is the symmetric QKD key. This ensures a cryptographically sound method of ``binding'' the different keys and \SecState together (which will be argued for in the security proof). Now, the protocol enters the authentication stage, beginning with the authentication of the responder.

The responder computes the encryption of $cert_R$ using the encryption algorithm of the Authenticated Encryption with Associated Data (AEAD) scheme with encryption key $tk_{shs}$ (and with the associated data string “Message 3”) and sends the resulting ciphertext $m_3$ to the initiator. Moreover, the responder computes the signature $\delta_R$ using the Digital Signature Scheme (DSS) signing algorithm with the signing key $\sk_R$ and the message $l_13||H_2$, and sends $\delta_R$ (AEAD-encrypted under $tk_{shs}$) to the initiator. (Note that $\sk_R$ may be an empty string, in which case DSS authentication is not used. In such situations, pre-shared key authentication must be in place, as described in the following section.)

The responder then computes the authentication tag $\tau_R$ using the MAC authentication algorithms with the key $\prfeval(\psk,fk_S)$, and sends it to the initiator. Here, the pre-shared key is used for authenticating the responder. (Note that \psk may be an empty string. In this case, pre-shared key authentication must not be used, and DSS authentication must therefore be in place. If \psk is an empty or known string, key confirmation can still be achieved via $fk_S$.)

The initiator now computes the decryptions of the received ciphertexts $m_3, m_4$ and $m_5$ using the AEAD decryption algorithm with decryption key $tk_{shs}$. The initiator verifies the certificate $cert_R$ (the exact method of which is beyond the scope of this work), the signature $\delta_R$, and the MAC tag $\tau_R$.

Now, both entities can derive the same master secret $MS$, and can update the secure state \SecState. This is followed by a authentication/confirmation step of the initiator, which proceeds analogously to the responder authentication described above. If all checks pass, this stage of the protocol is completed. The secure state \SecState derived during the completed stage is given as input for the next stage.

\begin{figure}[h]
\centering
\resizebox{\textwidth/2}{!}{
\begin{tikzpicture}[->,-latex,shorten >=1pt,auto,node distance=25mm, font=\Large,
  every state/.style={rectangle, rounded corners = 0.5mm},
  accepting/.style ={draw=none, fill=none},
  initial/.style ={draw=none, text=white}]
    \node[state]                     (0)                      {$k_0$};
    \node[state]                     (1)  [above right of = 0]{$k_1$};
    \node[state]                     (2)  [above right of = 1]{$k_2$};
    \node[state]                     (3)  [above right of = 2]{$k_3$};
    \node[state]                     (4)  [above left  of = 0]{$k_{pq}$};
    \node[state]                     (5)  [above left  of = 1]{$k_c$};
    \node[state]                     (6)  [above left  of = 3]{$SecState$};
    \node[state]                     (7)  [below right of = 3]{$dHS$};
    \node[state,initial,fill=clrQKD] (a)  [above left  of = 2]{$k_q$};
    \node[state,initial,fill=clrKPQ] (b)  [above left  of = 4]{$ss_{pq}$};
    \node[state,initial,fill=clrKC]  (c)  [above left  of = 5]{$ss_c$};
    \node[state,initial,fill=black]  (d)  [below right of = 7]{$MS$};
    \node[state]                     (8)  [below right of = d]{$SecState$};

    \path (0)  edge node {}(1)
          (1)  edge node {}(2)
          (2)  edge node {}(3)
          (3)  edge node {}(7)
          (4)  edge node {}(0)
          (5)  edge node {}(1)
          (6)  edge node {}(3)
          (7)  edge node {}(d)
          (a)  edge node {}(2)
          (b)  edge node {}(4)
          (c)  edge node {}(5)
          (d)  edge node {}(8);
          
    \draw[dashed]   (6) ++(-1.25,1.25) -- (6);
    \draw[dashed,-] (8) -- +(1.25,-1.25);
    \draw[clrQKD,font=\small]  (a)++(-2cm,0) -- (a)         node[above,midway]{$\GetKey_{qkd}$};
    \draw[clrKPQ,font=\small]  (b)++(-2cm,0) -- (b)         node[above,midway]{$\KEM_{pq}$};
    \draw[clrKC,font=\small]   (c)++(-2cm,0) -- (c)         node[above,midway]{$\KEM_c$};
    \draw (6.east) ++(4,0.5) coordinate(prf) -- +(0.5,-0.5) node[above right]{PRF};
    \draw (prf) ++(0,-1.25)  coordinate(drf) -- +(0.5,0.5)  node[below right]{dual-PRF};
\end{tikzpicture}
}
\caption {The key schedule of one stage of the \VMuckle protocol, illustrating how the key components are securely combined to create the key values $MS$ and $SecState$.}
\label{fig:key-schedule}
\end{figure}
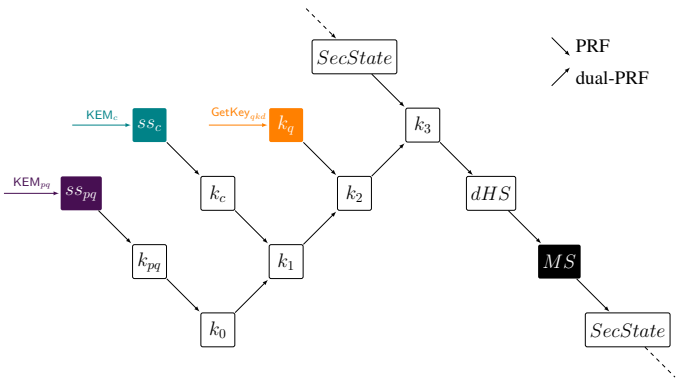

The key schedule of the \VMuckle stage is given in \Cref{fig:key-schedule}. Note that if at least one of the authentication mechanisms (DSS or PSK) can be verified successfully on both sides, correctness follows (under the assumptions of the correctness of the KEM building blocks and the QKD key derivation).

We provide the security analysis of \VMuckle in Appendix \ref{app:proof}.

\section{Evaluation of the \VMuckle Protocol}

\updated{To assess the practical performance of \VMuckle, we developed a prototypical implementation of the \VMuckle protocol. This prototype was created in Python, utilizing liboqs~\cite{SteMos16} bindings for post-quantum signature schemes and key encapsulation mechanisms, while also supporting conventional cryptography modules\footnote{\url{https://pypi.org/project/cryptography/}}. The protocol was evaluated on a notebook running Ubuntu 22.04.5 with an Intel(R) Core(TM)~i7-8665U CPU @ 1.90GHz and 16~GB of RAM.}

\updated{Simulators were employed instead of actual QKD devices to prevent the limited key rate from affecting the practical results. The benchmark involved mutual authentication, meaning both parties authenticated themselves using certificates. Similarly to the evaluation of Muckle+ in \cite{BruRamStr23} and \cite{battarbee2024quantumsafehybridkeyexchanges}, the certificates included both post-quantum and classical long-term public keys. They were authenticated within a two-layer certificate hierarchy, consisting of one root CA and one intermediate CA. The CAs signed the certificates using both a post-quantum signature scheme (ML-DSA-87) and a classical signature scheme, specifically EdDSA~\cite{BDLSY11}.}


\updated{The results are shown in Tables \ref{tab:performance-1} and \ref{tab:performance-2}.}
\begin{table}[htbp]
\centering
\caption{Bandwidth and CPU usage (part 1) in kilobytes and giga cycles, respectively for encapsulation (NIST level 1, 3, or 5) and signature (NIST levels 1, 3, or 5) variants in VMuckle. For QKD we have used QKD-256 bits. $^{(2)}$ For ECDH we have used ECDH-P521.}
\label{tab:performance-1}
\begin{tabular}{llcc}
\toprule
KEM & Signature & Cycles (G) & Data (KB) \\
\midrule
\multirow{9}{*}{\shortstack{ML-KEM-512 \\\& QKD$^{(1)}$\\\& ECDH$^{(2)}$}} 
  & PSK & <1.5 & 1.0 \\
  & ML-DSA-44 & 1.5 & 16.9 \\
  & ML-DSA-65 & 1.5 & 18.4 \\
  & ML-DSA-87 & 1.5 & 20.3 \\
  & SLH-DSA-SHAKE-128f & 1.6 & 30.3 \\
  & SLH-DSA-SHAKE-192f & 1.7 & 48.9 \\
  & SLH-DSA-SHAKE-256f & 1.9 & 63.1 \\
  & Falcon-512 & 1.6 & 14.7 \\
  & Falcon-1024 & 1.6 & 16.2 \\
\midrule
\multirow{9}{*}{\shortstack{ML-KEM-768 \\\& QKD$^{(1)}$\\\& ECDH$^{(2)}$}} 
  & PSK & <1.5 & 1.4 \\
  & ML-DSA-44 & 1.5 & 17.3 \\
  & ML-DSA-65 & 1.5 & 18.8 \\
  & ML-DSA-87 & 1.5 & 20.7 \\
  & SLH-DSA-SHAKE-128f & 1.8 & 30.7 \\
  & SLH-DSA-SHAKE-192f & 1.8 & 49.3 \\
  & SLH-DSA-SHAKE-256f & 1.9 & 63.5 \\
  & Falcon-512 & 1.5 & 15.1 \\
  & Falcon-1024 & 1.6 & 16.6 \\
\midrule
\multirow{9}{*}{\shortstack{ML-KEM-1024 \\\& QKD$^{(1)}$\\\& ECDH$^{(2)}$}} 
  & PSK & <1.5 & 1.8 \\
  & ML-DSA-44 & 1.5 & 17.7 \\
  & ML-DSA-65 & 1.5 & 19.2 \\
  & ML-DSA-87 & 1.5 & 21.1 \\
  & SLH-DSA-SHAKE-128f & 1.6 & 31.1 \\
  & SLH-DSA-SHAKE-192f & 1.7 & 49.7 \\
  & SLH-DSA-SHAKE-256f & 2.2 & 63.8 \\
  & Falcon-512 & 1.7 & 15.5 \\
  & Falcon-1024 & 1.7 & 17.0 \\
\midrule
\multirow{9}{*}{\shortstack{HQC-128 \\\& QKD$^{(1)}$\\\& ECDH$^{(2)}$}} 
  & PSK & <1.5 & 2.5 \\
  & ML-DSA-44 & 1.5 & 18.4 \\
  & ML-DSA-65 & 1.5 & 19.9 \\
  & ML-DSA-87 & 1.5 & 21.8 \\
  & SLH-DSA-SHAKE-128f & 1.7 & 31.2 \\
  & SLH-DSA-SHAKE-192f & 1.9 & 50.3 \\
  & SLH-DSA-SHAKE-256f & 1.9 & 64.5 \\
  & Falcon-512 & 1.6 & 16.2 \\
  & Falcon-1024 & 1.7 & 17.7 \\
\midrule
\multirow{9}{*}{\shortstack{HQC-192 \\\& QKD$^{(1)}$\\\& ECDH$^{(2)}$}} 
  & PSK & <1.6 & 4.8 \\
  & ML-DSA-44 & 1.6 & 20.6 \\
  & ML-DSA-65 & 1.6 & 22.1 \\
  & ML-DSA-87 & 1.8 & 24.1 \\
  & SLH-DSA-SHAKE-128f & 1.7 & 34.0 \\
  & SLH-DSA-SHAKE-192f & 1.9 & 52.6 \\
  & SLH-DSA-SHAKE-256f & 2.1 & 66.8 \\
  & Falcon-512 & 1.8 & 18.4 \\
  & Falcon-1024 & 1.8 & 20.0 \\
\bottomrule
\end{tabular}
\end{table}

\begin{table}[htbp]
\centering
\caption{Continuation of table \ref{tab:performance-1}.}
\label{tab:performance-2}
\resizebox{\columnwidth}{!}{
\renewcommand{\thetable}{3 -- continuation}
\begin{tabular}{llcc}
\toprule
KEM & Signature & Cycles (G) & Data (KB) \\
\midrule
\multirow{9}{*}{\shortstack{HQC-256 \\\& QKD$^{(1)}$\\\& ECDH$^{(2)}$}} 
  & PSK & <1.7 & 7.5 \\
  & ML-DSA-44 & 1.9 & 23.4 \\
  & ML-DSA-65 & 1.7 & 24.9 \\
  & ML-DSA-87 & 1.8 & 26.8 \\
  & SLH-DSA-SHAKE-128f & 1.9 & 36.7 \\
  & SLH-DSA-SHAKE-192f & 1.9 & 55.3 \\
  & SLH-DSA-SHAKE-256f & 2.1 & 69.5 \\
  & Falcon-512 & 1.8 & 21.2 \\
  & Falcon-1024 & 2.0 & 22.7 \\
\midrule
\multirow{9}{*}{\shortstack{FrodoKEM-640-SHAKE\\\& QKD$^{(1)}$\\\& ECDH$^{(2)}$}} 
  & PSK & <1.6 & 9.8 \\
  & ML-DSA-44 & 1.6 & 25.7 \\
  & ML-DSA-65 & 1.6 & 27.2 \\
  & ML-DSA-87 & 1.6 & 29.2 \\
  & SLH-DSA-SHAKE-128f & 1.9 & 39.1 \\
  & SLH-DSA-SHAKE-192f & 1.9 & 57.7 \\
  & SLH-DSA-SHAKE-256f & 2.1 & 71.9 \\
  & Falcon-512 & 1.8 & 23.5 \\
  & Falcon-1024 & 1.9 & 25.1 \\
\midrule
\multirow{9}{*}{\shortstack{FrodoKEM-976-SHAKE\\\& QKD$^{(1)}$\\\& ECDH$^{(2)}$}} 
  & PSK & <1.6 & 15.9 \\
  & ML-DSA-44 & 1.6 & 31.7 \\
  & ML-DSA-65 & 1.6 & 33.3 \\
  & ML-DSA-87 & 1.6 & 35.2 \\
  & SLH-DSA-SHAKE-128f & 1.7 & 45.1 \\
  & SLH-DSA-SHAKE-192f & 1.9 & 63.7 \\
  & SLH-DSA-SHAKE-256f & 2.0 & 77.9 \\
  & Falcon-512 & 1.6 & 29.6 \\
  & Falcon-1024 & 1.7 & 31.1 \\
\midrule
\multirow{9}{*}{\shortstack{FrodoKEM-1344-SHAKE\\\& QKD$^{(1)}$\\\& ECDH$^{(2)}$}} 
  & PSK & <1.6 & 21.8 \\
  & ML-DSA-44 & 1.6 & 37.6 \\
  & ML-DSA-65 & 1.6 & 39.1 \\
  & ML-DSA-87 & 1.6 & 41.1 \\
  & SLH-DSA-SHAKE-128f & 1.7 & 51.0 \\
  & SLH-DSA-SHAKE-192f & 1.8 & 69.6 \\
  & SLH-DSA-SHAKE-256f & 2.0 & 83.8 \\
  & Falcon-512 & 1.8 & 35.4 \\
  & Falcon-1024 & 1.9 & 37.0 \\
\bottomrule
\end{tabular}
}
\end{table}

\updated{To conclude, ML-KEM-512, ML-KEM-768 or ML-KEM-1024 with ML-DSA-44, ML-DSA-65, or ML-DSA-87 are notable for their efficiency in terms of both bandwidth \textit{and} CPU usage (using Falcon-512 or Falcon-1024 instead of the ML-DSA variants could be a valid choice for bandwidth-constrained use cases). On the other hand, FrodoKEM-1344-SHAKE and SLH-DSA-SHAKE-806-256f provide enhanced security but require more bandwidth and computational resources. The selection of a scheme should be based on the specific needs of the application, balancing bandwidth efficiency against computational performance in post-quantum cryptographic contexts.}

\updated{Overall, using ML-KEM-1024 in \VMuckle with ML-DSA-87 together with conventional cryptography and QKD yields a valuable choice for the integration into MACsec balancing efficiency and security.}

\updated{The alternative to \VMuckle is the EAP-TLS protocol. To assess the overhead introduced by \VMuckle, we will compare its performance with EAP-TLS. In both cases, we will use a mutual authentication scenario (both ends of the connection are authenticated). EAP-TLS requires 14 messages to operate: 4 are specific to the EAP protocol and 10 to the TLS protocol.}

\updated{The first message is the EAP-Request Identity, sent from the authenticator server to the client, which contains approximately 30 B of information. The client replies with the EAP-Response Identity message, including its username, which typically amounts to around 25 B depending on its length. The server then sends the EAP-Request TLS Start (6 B) to initiate the TLS handshake. The client continues with the EAP-Response Client Hello message, which includes 78 B in addition to the key size of the public key sent, assuming that only the necessary key material for the selected asymmetric algorithm is transmitted. The server responds with a sequence of messages: the EAP-Request Server Hello (72 B plus the size of the public key or ciphertext), EAP-Request Certificate (22 B plus the certificate size), EAP-Request Encrypted Extensions (15 B), EAP-Request CertificateRequest (24 B), EAP-Request CertificateVerify (17 B plus the signature size), and EAP-Request Finished (61 B, assuming SHA-384 is used). The client then sends the EAP-Response Certificate (22 B plus the certificate size), followed by the EAP-Request CertificateVerify (17 B plus the signature size), and the EAP-Response Finished (61 B). Finally, the server completes the handshake by sending the EAP-Success message (4 B). Together, the base size of the handshake is 424 B, not including the additional data contributed by public keys, ciphertexts, certificates, and signatures \cite{RFC_8446, RFC_9190} (i.e, for ML-KEM-1024 and ML-DSA-87, the total data sent is 17,29 KB). In contrast, with \VMuckle, the total data sent for this ciphersuite is 42,2 KB, representing an increase of around 144,07\% (note that it is only during the first round, so it is almost irrelevant from a performance point of view). Although \VMuckle is less efficient in terms of the amount of data exchanged -- due to its distinct protocol flow designed to support hybrid signature and keys -- this design enables a rigorous security proof, which EAP-TLS currently lacks.}

\section{Quantum-Safe MACsec}

As mentioned in Background (\ref{subsec:MACsec}), MACsec is defined in two different standards: 802.1AE and 802.1X. 802.1AE defines the MACsec protocol, which describes the process of creating a secure layer-2 level session once the transmitting and receiving keys have been established, but the authentication and key agreement are out of the scope of this standard.

802.1X defines both authentication and key establishment for MACsec as independent modules, where MKA is in charge of discovering peers inside a MACsec CA, and negotiating, deriving, and distributing keys, but authentication is out of the scope of the MKA protocol, since MKA uses an authenticated master key as root key for its key hierarchy. This process is performed by establishing and maintaining a secure communication channel between the stations within an CA by deriving the encryption (KEK) and integrity (ICK) keys from this root key provided by the authentication. 802.1X encourages the use of the authentication method defined in the standard for MKA, which may be performed through EAP-TLS or PSKs, but it allows the use of other means of authentication as long as they have been proven to be secure. This modularity offers greater flexibility, allowing a network to deploy 802.1X for authentication without implementing MACsec, or to use alternative authentication methods while still utilizing MKA to distribute keys for secure network communications.

In this work, we leverage the modularity of 802.1X to introduce a novel authentication method to agree on a root key for the MKA protocol, without requiring any modifications to existing standards. As we have detailed in the previous section, \VMuckle is proven to be a secure quantum-safe authenticated method to obtain a symmetric key at both endpoints of connection. This approach provides MKA with a quantum-safe HAKE for each station within the CA, allowing the initiation of integrity and encryption key establishment. Furthermore, not modifying the MKA protocol defined in 802.1X means that this implementation benefits from the proven security of this standard, augmented by the security of hybrid quantum-safe authentication.

\begin{figure}[htbp!]
\centering
    \includegraphics[width=.97\linewidth]{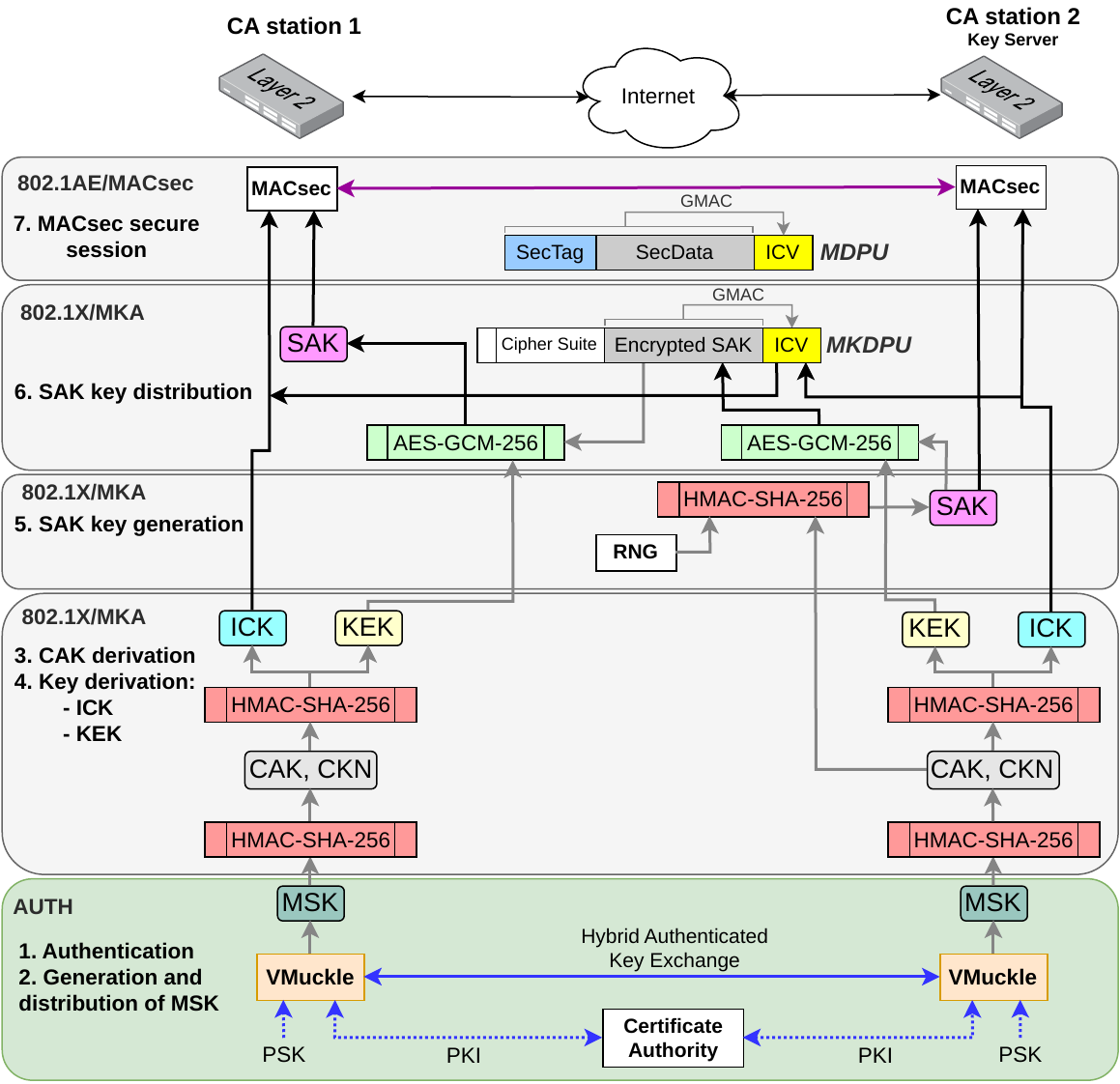}
\caption{The proposed operation for employing \VMuckle as source of trust for MKA protocol in a two-node scenario.}
\label{fig:Operation}
\end{figure}

The operation of the proposed implementation is illustrated in Figure \ref{fig:Operation}. The initial phase of the unmodified MACsec-MKA protocol flow involves authenticating connection endpoints using the authentication mechanism defined in 802.1X (depicted in green). In this work, we propose the use of \VMuckle to authenticate the endpoints and generate a root key (MSK) that serves as the source of the keying material for MKA. This phase is the only one that has been modified in the MACsec deployment operation flow.

From this point on, the operation of both the MACsec and MKA protocols proceeds without modification, ensuring the security guaranteed by both 802.1X and 802.1AE, using the quantum-safe hybrid key agreed by \VMuckle as the Master Session Key. The second phase consists of the generation of the Connectivity Association Key and its Connectivity Association Key Name for the MACsec Key Agreement protocol. The Connectivity Association Key Name contains the Connectivity Association Key identifier, configured manually or defined as a static value. The Connectivity Association Key is derived from the Master Session Key using HMAC over SHA-256, where the Master Session Key is used as the HMAC key and the Connectivity Association Key Name as the input message. Then, the Key Encryption Key and the Integrity Check Key are generated via HMAC over SHA-256 using the Connectivity Association Key as key and a specific derivation constraint (different for KEK and ICK) among with the Connectivity Association Key Name as input message.

Subsequently, a key server, located on the right side of Figure \ref{fig:Operation}, must be selected within the CA. The key server is responsible for deriving the Secure Association Key (shown in purple in Figure \ref{fig:Operation}) using the same KDF: HMAC over SHA-256, combining various parameters such as the Connectivity Association Key, the Secure Channel Identifier, and a synchronization value generated from an RNG.

The Secure Association Key then must be securely distributed to the other endpoint. To achieve this, the Secure Association Key is encrypted using AES-GCM-256, with the KEK serving as the encryption key, as specified in 802.1X. The encrypted Secure Association Key, among the Integrity Check Value (ICV) and the selected ciphersuite, is encapsulated within an MKDPU frame. This frame ensures both data integrity and confidentiality during transmission to the other endpoint of the Security Associations. Upon receiving the MKDPU frame, the participant device decrypts the Secure Association Key and validates the integrity of the frame. This process enables the secure establishment of the MACsec session, creating a secure communication channel between the participating SA devices by sending MACsec data protocol units (MDPU), as shown in the upper part of Figure \ref{fig:Operation}.

The primary advantage of our approach lies in its integration of post-quantum security through the combined use of classical cryptography, PQC, and QKD as a trust anchor within the 802.1X key management hierarchy. This architecture enables MACsec to establish a hybrid quantum-safe secure key agreement mechanism, which can be authenticated using PSKs, certificate-based methods, or a combination of both, providing a secure and scalable solution. As a result, this design ensures that MACsec benefits from post-quantum security without requiring modifications to existing standards, resulting in a smooth transition to quantum-safe networks with provable security guarantees on the authentication.

\section{Conclusion and Future Work}

In this work, we provide a new Hybrid Authenticated Key Exchange (HAKE) protocol dubbed \VMuckle, because of its versatility for large-- and small--scale quantum-safe networks, with features that were not available before. In addition, \VMuckle is used to obtain a secure authenticated symmetric key for MKA use between MACsec endpoints, ensuring robust quantum-safe security. Furthermore, integrating \VMuckle as a HAKE mechanism that can be readily used in the 802.1X standard facilitates a smooth transition towards quantum-safe infrastructures and ensures that MACsec remains a future-proof standard, capable of addressing both classical and quantum security threats.

Future work that we intend to tackle within this framework includes multicast capabilities. This would enable the generation of a CAK on the key server and its distribution to all the network nodes. While feasible from an MKA perspective, achieving this within the \VMuckle protocol presents challenges that require further investigations. In addition, to demonstrate the maturity of the solution, a logical next step would be to deploy the solution in a production setting. \updated{In this sense, the adoption of the protocol in a broader context would be improved by requiring the inclusion of an initial negotiation phase to check which services are available at the endpoint, so that other quantum-safe algorithms would be used in case QKD is unavailable. An additional line of work could be to extend the same authentication technologies to other protocols such as IPsec.}

Finally, we plan a full testing and deployment on the MadQCI network \cite{MadQCI}. This is a large and heterogeneous network that is currently being enlarged with connections to final users\updated{, getting closer to a real production network,} and that will serve as an ideal testbed for the work described here. 

\appendices

\section{Mathematical Preliminaries}\label{app:pre-hake}

\paragraph{Notation.} 
    Let $\secpar \in \N$ be the security parameter. For a finite set $\setS$, we denote by $s\gets\setS$ the process of sampling an element $s$ uniformly from $\setS$. For an algorithm $A$, we let $y \gets A(\secpar,x)$ denote the process of running $A$ on input $(\secpar,x)$ with access to uniformly random coins, and assigning the result to $y$. (We may omit explicit mention of the $\secpar$-input and assume that all algorithms take $\secpar$ as input.) We say an algorithm $A$ is probabilistic polynomial time (PPT) if the running time of $A$ is polynomial in $\secpar$ \updated{ by a probabilistic Turing machine. An algorithm $A$ is called quantum polynomial time (QPT) if it is a uniform family of quantum circuits with size polynomial in $\secpar$}. A function $f$ is called negligible if its absolute value is smaller than the inverse of any polynomial, for sufficiently large input values (i.e., if $\forall c\in\N\ \exists k_0\ \forall \kk \geq k_0:|f(\kk)|<1/ \kk^c$). 

\updated{In the following, we recap standard (classical) notions of the cryptographic building blocks needed. At the end of the section, we note that post-quantum equivalents can also be similarly formulated.}

\begin{definition}[Pseudo-Random Function]
    Let $\prfeval\colon \mathcal{S} \times D \to \mathsf{R}$ be a family of functions, and let $\Gamma$ be the set of all functions $D \rightarrow \mathsf{R}$. For a PPT distinguisher $\mathcal{D}$ we define the advantage function as $\Adv^{\mathsf{PRF}}_{\mathcal{D},\prfeval}(\secpar) =$
    \begingroup
        \fontsize{8.9pt}{12pt}
        \selectfont
        \begin{align*}
            \left| \Pr\left[s \gets \mathcal{S}: {\cal D}^{\prfeval(s, \cdot)}(\secpar) = 1\right] - \Pr\left[f\gets \Gamma: {\cal D}^{f(\cdot)}(\secpar) = 1\right] \right| \text{.}
        \end{align*}
    \endgroup
    $\prfeval$ is called a pseudorandom function (family) if it is efficiently computable and, for any PPT distinguisher $\mathcal{D}$, there exists a negligible function $\varepsilon(\cdot)$ such that
    \[
        \Adv^{\mathsf{prf}}_{\mathcal{D},\prfeval}(\secpar) \leq \varepsilon(\secpar) \text{.}
    \]
    A PRF $\prfeval$ is called a dual PRF~\cite{EPRINT:BelLys15}, if $\mathcal{G} \colon D \times \mathcal{S} \to \mathsf{R}$ defined by $\mathcal{G}(d, s) = \prfeval(s, d)$ is also a PRF.
\end{definition}

We recall the notion of message authentication codes (MACs) as well as digital signature schemes, and the standard unforgeability notions below.

\begin{definition}[Message Authentication Codes] \label{def:macs}
    A message authentication code $\MAC$ is a triple $(\kgen,\allowbreak \ssign, \sverify)$ of PPT algorithms, which are defined as:
    \begin{description}
        \item[$\kgen(\secpar)\colon$] This algorithm takes a security parameter $\secpar$ as input, and outputs a secret key $\sk$.
        \item[$\msign(\sk, \msg)\colon$] This algorithm takes a secret key $\sk \in \Key$ and a message $\msg \in \Msg$ as input, and outputs an authentication tag $\tau$.
        \item[$\mverify(\sk,\msg,\tau)\colon$] This algorithm takes a secret key $\sk$, a message $\msg \in \Msg$ and an authentication tag $\tau$ as input, and outputs a bit $b \in \{0,1\}$.
    \end{description}
\end{definition}

A MAC is correct if, for all $\secpar \in \N$, for all $\sk \gets \kgen(\secpar)$ and for all $\msg \in \Msg$, it holds that
    \[
        \Pr\left[ \mverify(\sk, \msg, \allowbreak\msign(\sk,\msg))=1 \right] =1 \text{,} 
    \]
where the probability is taken over the random coins of $\kgen$ and $\msign$.

\begin{definition}[$\EUFCMA$ security of \MAC]
    For a PPT adversary $\advA$, we define the advantage function in the sense of existential unforgeability under chosen message attacks ($\EUFCMA$) as
    \[
        \AdvEUFCMA{\advA,\MAC}(\secpar) = \Pr\left[ \ExpEUFCMA{\advA,\MAC}(\secpar) = 1\right] \text{,}
    \]
    with the corresponding experiment depicted in \Cref{fig:mac-unfcma}. If, for all PPT adversaries $\adv$, there is a negligible function $\varepsilon(\cdot)$ such that
    \(
        \AdvEUFCMA{\advA,\MAC}(\secpar) \leq \varepsilon(\secpar) \text{,}
    \)
    we say that $\MAC$ is $\EUFCMA$ secure.
\end{definition}

\begin{experiment}[ht]
    \centering
    \pseudocode[mode=text]{
        $\ExpEUFCMA{\advA,\MAC}(\secpar)$: \\
        \t $\sk \gets \kgen(\secpar)$, $\mathcal{Q} \gets \emptyset$ \\
        \t $(\msg^*, \tau^*)\gets\advA^{\msign',\mverify'}(\secpar)$ \\
        \t \t where oracle $\msign'(\msg)$: \\
        \t \t \t $\mathcal{Q} \gets \mathcal{Q} \cup \{\msg \}$ \\
        \t \t \t return $\msign(\sk, \msg)$ \\
        \t \t where oracle $\mverify'(\msg, \tau)$: \\
        \t \t \t return $\mverify(\sk, \msg, \tau)$ \\
        \t return $1$ if $\mverify(\sk, \msg^*, \tau^*) = 1~\land~\msg^*\notin \mathcal{Q}$ return $0$, otherwise \\
    }
\caption{$\EUFCMA$ security experiment for a MAC $\MAC$.}
\label{fig:mac-unfcma}
\end{experiment}

\begin{definition}[Signature Scheme] \label{def:sigs}
    A signature scheme $\DSIG$ is a triple $(\kgen,\allowbreak \ssign, \sverify)$ of PPT algorithms, which are defined as follows:
    \begin{description}
        \item[$\kgen(\secpar)\colon$] This algorithm takes a security parameter $\secpar$ as input and outputs a secret (signing) key $\sk$ and a public (verification) key $\pk$ with associated message space $\Msg$ (we may omit to make the message space $\Msg$ explicit).
        \item[$\ssign(\sk, \msg)\colon$] This algorithm takes a secret key $\sk$ and a message $\msg \in \Msg$ as input, and outputs a signature $\sigma$.
        \item[$\sverify(\pk,\msg,\sigma)\colon$] This algorithm takes a public key $\pk$, a message $\msg \in \Msg$ and a signature $\sigma$ as input, and outputs a bit $b \in \{0,1\}$.
    \end{description}
\end{definition}
For correctness, we require that for all $\secpar\in \mathbb{N}$, for all $(\sk, \pk) \gets \kgen(\secpar)$ and for all $\msg \in\Msg$ it holds that
\[
    \Pr\left[ \sverify(\pk, \msg, \ssign(\sk,\msg))=1 \right] =1 \text{,}
\]
where the probability is taken over the random coins of $\kgen$ and $\ssign$.

\begin{definition}[$\EUFCMA$ of \DSIG]
    For a PPT adversary $\advA$, we define the advantage function in the sense of existential unforgeability under chosen message attacks ($\EUFCMA$) as
    \[
        \AdvEUFCMA{\advA,\DSIG}(\secpar) = \Pr\left[ \ExpEUFCMA{\advA,\DSIG}(\secpar) = 1\right] \text{,}
    \]
    where the corresponding experiment is depicted in \Cref{fig:dsig-euf-cma}. If for all PPT adversaries $\adv$ there is a negligible function $\varepsilon(\cdot)$ such that
    \(
        \AdvEUFCMA{\advA,\DSIG}(\secpar) \leq \varepsilon(\secpar) \text{,}
    \)
    we say that $\DSIG$ is $\EUFCMA$ secure.
\end{definition}

\begin{experiment}[ht]
    \centering
    \pseudocode[mode=text]{
        $\ExpEUFCMA{\advA,\DSIG}(\secpar)$: \\
        \t $(\sk, \pk) \gets \kgen(\secpar)$, $\mathcal{Q} \gets \emptyset$ \\
        \t $(\msg^*, \sigma^*)\gets\advA^{\ssign}(\pk)$ \\
        \t \t where oracle $\ssign'(\msg)$: \\
        \t \t \t $\mathcal{Q} \gets \mathcal{Q} \cup \{\msg \}$ \\
        \t \t \t return $\ssign(\sk, \msg)$ \\
        \t return $1$ if $\sverify(\pk, \msg^*, \sigma^*) = 1~\land~\msg^*\notin \mathcal{Q}$, return $0$ otherwise \\
    }
\caption{$\EUFCMA$ security experiment for a digital signature scheme $\DSIG$.}
\label{fig:dsig-euf-cma}
\end{experiment}

We recall the notion of key-encapsulations mechanisms (KEMs) and the standard chosen-plaintext notion below.
\begin{definition}[Key-Encapsulation Mechanism]
    A key-encapsulation mechanism scheme $\KEM$ with key space $\Key$ consists of the three PPT algorithms $(\kgen,\Encaps,\Decaps)$:
    \begin{description}
        \item[$\kgen(\secpar)\colon$] This algorithm takes a security parameter $\secpar$ as input, and outputs public and secret keys $(\pk,\sk)$.
        \item[$\Encaps(\pk)\colon$] This algorithm takes a public key $\pk$ as input, and outputs a ciphertext $\ctxt$ and a key $\key$.
        \item[$\Decaps(\sk, \ctxt)\colon$] This algorithm takes a secret key $\sk$ and a ciphertext $\ctxt$ as input, and outputs $\key$ or $\{\bot\}$.
    \end{description}
\end{definition}

We call a $\KEM$ correct if, for all $\secpar\in\N$, for all $(\pk,\sk)\gets\kgen(\secpar)$, and for all $(\ctxt,\key)\gets\Encaps(\pk)$, we have that
\[
  \Pr[\Decaps(\sk,\ctxt)=\key] = 1 \text{,}
\] 
where the probability is taken over the random coins of $\kgen$ and $\Encaps$.

\begin{definition}[IND-CPA security of \KEM]
    For a PPT adversary $\advA$, we define the advantage functions in the sense of indistinguishability under chosen-plaintext attacks (\KEMINDCPA) as
    \begin{align*}
        \AdvINDCPA{\advA,\KEM}(\secpar) &= \left| \Pr\left[ \ExpKEMINDCPA{\advA,\KEM}(\secpar) = 1\right] -\frac{1}{2} \right|
    \end{align*}
    where the corresponding experiment is depicted in \Cref{fig:kem-ind-cpa}. If for all PPT adversaries $\adv$ there is a negligible function $\varepsilon(\cdot)$ such that
    \[
        \AdvINDCPA{\advA,\KEM}(\secpar) \leq \varepsilon(\secpar) \text{,}
    \]
    then we say that $\KEM$ is \KEMINDCPA secure.
\end{definition}

\begin{experiment}[t]
    \centering
    \pseudocode[mode=text]{
    $\ExpINDCPA{\advA,\KEM}(\secpar)$: \\
    \t $(\pk, \sk) \gets \kgen(\secpar)$ \\
    \t $(\ctxt^*,\key_0)\gets\Encaps(\pk),\key_1\gets\Key$ \\
    \t $b\gets\{0,1\}^\secpar$ \\
    \t $b^* \gets\advA(\pk, \ctxt^*, \key_b)$ \\
    \t return $1$, if $b = b^*$, return $0$ otherwise \\
    }
\caption{IND-CPA security experiment for a KEM \KEM.}
\label{fig:kem-ind-cpa}
\end{experiment}

\updated{
\begin{note}[Post-Quantum Security]
    The security notions introduced above describe classical security (i.e., security against PPT adversaries). We would also like to be able to discuss security against quantum capable adversaries. For each security definition, we define the corresponding "post-quantum secure" equivalent by requiring that the advantage bound also holds against all QPT adversaries.
\end{note}
}

We further use the Authenticated Encryption with Associated Data (AEAD) scheme as defined in \cite{RBBK01}.

\section{Hybrid Authenticated Key Exchange}

We recall the hybrid authenticated key exchange (HAKE) security model~\cite{DowHanPat20,BruRamStr23}. The HAKE security experiment $\Exp^{\mathsf{hake},\clean}_{\advA,\Pi,n_P,n_S,n_T}$ is described as in \cite[Fig. 5, App. C]{DowHanPat20}. Here, we only recall the execution environment, adversarial interaction, and matching sessions.

\paragraph{Execution environment.}
We consider a set of $n_P$ parties, $P_1, \ldots, P_{n_P}$, each of which is able to run up to $n_S$ sessions of a key-exchange protocol $\Pi$. A session consists of up to $n_T$ stages of the protocol. Each party $P_i$ has access to its long-term key pair $(\pk_i, \sk_i)$ and to the public keys of all other parties. Every session is described by a set of session parameters:
\begin{itemize}
    \item $\rho \in \{\mathsf{init},\mathsf{resp}\}$: The role (initiator or responder) of the party during the current session.
    \item $pid \in n_P$: The communication partner of the current session.
    \item $stid \in n_T$: The current stage of the session.
    \item $\alpha \in \{\act,\acc,\rej,\bot\}$: The status of the session. Initialized with $\bot$.
    \item $m_i[stid], i \in \{s,r\}$: All messages sent ($i=s$) or received ($i=r$) by a session up to the stage $stid$. Initialized with $\bot$.
    \item $k[stid]$: All session keys created up to stage $stid$. Initialized with $\bot$.
    \item $exk[stid], x \in \{q,c,s\}$: All ephemeral post-quantum (q), classical (c) or symmetric (s) secret keys created up to stage $stid$. Initialized with $\bot$.
    \item $pss[stid]$: The per-session secret state (\SecState) that is created during the stage $stid$ for use in the next stage.
    \item $st[stid]$: Storage for other states used by the session in each stage.
\end{itemize}
The $s$-th instance of the protocol run by a given party $P_i$ is denoted by $\pi_i^s$, and the value of a particular parameter $\mathsf{par}$ is referenced using $\pi_i^s.\mathsf{par}$.

We describe the protocol as a set of algorithms $(f$, $\kgen{}XY$, $\kgen{}ZS$):
\begin{itemize}
    \item $f(\secpar, \pk_i, \sk_i, pskid_i, \psk_i,\pi,m) \rightarrow (m',\pi')$: A probabilistic algorithm that represents an honest execution of the protocol. It takes a security parameter $\secpar$, the long-term keys $(\pk_i, \sk_i)$, the session parameters $\pi$ representing the current state of the session, and a message $m$ as input, and outputs a response $m'$ and the updated session state $\pi'$.
    \item $\kgen{}XY(\secpar) \rightarrow (\pk,\sk)$: A probabilistic asymmetric key-generation algorithm that takes a security parameter $\secpar$ and creates a public-secret-key pair $(\pk,\sk)$. $X \in \{E,L\}$ determines whether the created key is an ephemeral (E) or a long-term (L) secret key. $Y \in \{Q,C\}$ determines whether the key is classical (C) or post-quantum (Q).
    \item $\kgen{}ZS(\secpar) \rightarrow (\psk,pskid)$: A probabilistic symmetric key-generation algorithm that takes a security parameter $\secpar$ and outputs symmetric keying material $(\psk)$. $Z \in \{E,L\}$ determines whether the created key is an ephemeral (E) or a long-term (L) secret key.
\end{itemize}

For each party $P_1, \ldots, P_{n_P}$, classical as well as post-quantum long-term keys are created using the corresponding $\kgen{}XY$ algorithms. The challenger then queries a uniformly random bit $b \gets \{0,1\}$, which will determine the key returned by the $\Test$ query. From this point on, the adversary may interact with the challenger using the queries defined in the next section. At some point during the execution of the protocol, the adversary $\advA$ may issue the $\Test$ query and present a guess for the value of $b$. If $\advA$ guesses correctly and the session satisfies the cleanness predicate, the adversary wins the key-indistinguishability experiment.

\paragraph{Adversarial Interaction.}
The HAKE framework defines a range of queries that allow the attacker to interact with the communication:
\begin{itemize}
    \item $\Create(i,j,role) \rightarrow \{(s),\bot\}$: Initializes a new session between the party $P_i$ with role $role$ and the partner $P_j$. If the session already exists, then the query returns $\bot$, otherwise the session $(s)$ is returned.
    \item $\Send(i,s,m) \rightarrow \{m',\bot\}$: Enables $\advA$ to send messages to sessions and receive the response $m'$ by running $f$ for session $\pi_i^s$. Returns $\bot$ if the session is not active.
    \item $\Reveal(i,s,t)$: Provides $\advA$ with the session keys corresponding to a session $\pi_i^s$ if the session is in the accepted state. Otherwise, $\bot$ is returned.
    \item $\Test(i,s,t) \rightarrow \{k_b,\bot\}$: Provides $\advA$ with the real (if $b=1$) or random ($b=0$) session key for the key-indistinguishably experiment.
    \item $\Corrupt{}XY(i) \rightarrow \{key,\bot\}$: Provides $\advA$ with the long-term $XY \in \{\mathsf{SK},\mathsf{QK},\mathsf{CK}\}$ keys for $P_i$. If the key has been previously corrupted, then $\bot$ is returned. Specifically:
    \begin{itemize}
        \item $\CorruptSK$: Reveals the long-term symmetric secret, i.e., the pre-shared key (if available).
        \item $\CorruptQK$: Reveals the post-quantum long-term key (if available).
        \item $\CorruptCK$: Reveals the classical long-term key (if available).
    \end{itemize}
    \item $\Compromise{}XY(i,s,t) \rightarrow \{key,\bot\}$: Provides $\advA$ with the ephemeral $XY \in \{ \mathsf{QK}, \mathsf{CK}, \mathsf{SK}, \mathsf{SS} \}$ keys created during the session $\pi_i^s$ prior to stage $t$. If the ephemeral key has already been compromised, then $\bot$ is returned. Specifically:
    \begin{itemize}
        \item $\CompromiseQK$: Reveals the ephemeral post-quantum key.
        \item $\CompromiseCK$: Reveals the ephemeral classical key.
        \item $\CompromiseSK$: Reveals the ephemeral quantum key.
        \item $\CompromiseSS$: Reveals the ephemeral per session state (\SecState).
    \end{itemize}
\end{itemize}

\paragraph{Matching sessions.}
Furthermore, we recall the definitions of matching sessions~\cite{LKZC07} and origin sessions~\cite{CreFel12} which covers that the two parties involved in a session have the same view of their conversation.

\begin{definition}[Matching sessions]
    We consider two sessions $\pi_i^s$ and $\pi_j^r$ in stage t to be matching if all messages sent by the former session $\pi_i^s.m_s[t]$ match those received by the later $\pi_j^r.m_r[t]$ and all messages sent by the later session $\pi_j^r.m_s[t]$ are received by the former $\pi_i^s.m_r[t]$.

    $\pi_i^s$ is considered to be prefix-matching with $\pi_j^r$ if $\pi_i^s.m_s[t]$ $= \pi_j^r.m_r[t]'$ where $\pi_j^r.m_r[t]$ is truncated to the length of $\pi_i^s.m_s[t]$ resulting in $\pi_j^r.m_r[t]'$.
\end{definition}

\begin{definition}[Origin sessions]
    We consider a session $\pi_i^s$ to have an origin session with $\pi_j^r$ if $\pi_i^s$ matches $\pi_j^r$ or if $\pi_i^s$ prefix-matches $\pi_j^r$.
\end{definition}

\begin{definition}[HAKE security]
    Let $\Pi$ be a key-exchange protocol, and $n_P, n_S, n_T \in \N$. For a predicate $\clean$ and an adversary $\advA$, we define the advantage of $\advA$ in the HAKE key-indistinguishability game as
    \[
        \Adv^{\mathsf{hake},\clean}_{\advA,\Pi,n_P,n_S,n_T}(\secpar) = \left| \Pr\left[ \Exp^{\mathsf{hake},\clean}_{\advA,\Pi,n_P,n_S,n_T}(\secpar) = 1 \right] \right|\text{.}
    \]
    We say that $\Pi$ is HAKE-secure if $\Adv^{\mathsf{hake},\clean}_{\advA,\Pi,n_P,n_S,n_T}(\secpar)$ is negligible in the security parameter $\secpar$ for \updated{any PPT adversary} $\advA$. \updated{We say $\Pi$ is post-quantum HAKE-secure if the advantage is also negligible against any QPT adversary.}
\end{definition}

\section{Security of \VMuckle}\label{app:proof}

We define two new cleanness predicates, $\clean_\VM$ and $\clean_{c\VM}$, and determine the conditions under which we have the desired security properties (post-compromise security, perfect forward secrecy). 

\begin{definition}[Cleanness Predicate]
    A session $\pi_i^s$ in stage $t$ is considered clean under the predicate $\clean_{\VM}$ if:  
    \begin{enumerate}
        \item $\Reveal(i,s,t)$ has not been issued for session $\pi_i^s$.
        \item $\Reveal(j,r,t)$ has not been issued for any session $\pi_j^r$ matching $\pi_i^s$ in stage $t$.
        \item If $\pi_i^s$ has a matching session $\pi_j^r$, at least one of the following holds:
        \begin{enumerate}[label=\roman*., align=left]
            \item No $\CompromiseQK(i,s,t)$ or $\CompromiseQK(j,r,t)$ have been issued.
            \item No $\CompromiseSK(i,s,t)$ or $\CompromiseSK(j,r,t)$ have been issued.
            \item No $\CompromiseQK(i,s,t')$ or $\CompromiseQK(j,r,t')$ have been issued with $\pi^s_i$ matching $\pi^r_j$ in stages $u$ where $t' \leq u \leq t$. No $\CompromiseSS(i,s,u)$, $\CompromiseSS(j,r,u)$ have been issued (except possibly in stage $t'$). No $\Reveal(i,s,u)$ or $\Reveal(j,r,u)$ have been issued.
            \item No $\CompromiseSK(i, s, t')$ or $\CompromiseSK(j, r, t')$ have been issued with $\pi^s_i$ matching $\pi^r_j$ in stages $u$ where $t' \leq u \leq t$. No $\CompromiseSS(i,s,u)$ or $\CompromiseSS(j, r, u)$ have been issued (except possibly in stage $t'$). No $\Reveal(i,s,u)$ or $\Reveal(j,r,u)$ have been issued.
        \end{enumerate}
        \item Either no $\CorruptQK(i)$ or no $\CorruptSK$ queries have been issued before $\pi_i^s.[t]\leftarrow accept$.
        \item If $\pi_j^r$ is an origin session of $\pi_i^s$ in stage $t$, then either no $\CorruptQK(j)$ or no $\CorruptSK$ queries have been issued before $\pi_j^r.[t]\leftarrow accept$.
    \end{enumerate}
    The predicate $\clean_{c\VM}$ differs from $\clean_{\VM}$ by the inclusion of two additional alternatives in condition 3:
    \begin{itemize}[label={}]
    \item
        \begin{enumerate}[label=\roman*., align=left]
        \setcounter{enumi}{4}
            \item No $\CompromiseCK(i,s,t)$ or $\CompromiseCK(j,r,t)$ have been issued.
            \item No $\CompromiseCK(i, s, t')$ or $\CompromiseCK(j, r, t')$ have been issued with $\pi^s_i$ matching $\pi^r_j$ in stages $u$ where $t' \leq u \leq t$. No $\CompromiseSS(i,s,u)$ or $\CompromiseSS(j, r, u)$ have been issued (except possibly in stage $t'$). No $\Reveal(i,s,u)$ or $\Reveal(j,r,u)$ have been issued.
        \end{enumerate}
    \end{itemize}
\end{definition}

The first and second conditions ensure that the session key of the session used in the $\Test$ query has not been directly revealed to the adversary through the use of the $\Reveal$ query. The third condition guarantees that at least one ephemeral secret has not been compromised, or in the multi-stage setting, that the secret state has not been compromised. In particular, case (i) covers the situation where the post-quantum key has not been compromised, case (ii) applies when the quantum key has not been compromised, and cases (iii) and (iv) cover the scenario in which a previous stage was completed cleanly and the session key has not been compromised for any intermediate stage. \updated{Cases (v) and (vi) are the classical analogues of (i) and (iii).} The fourth and fifth conditions specify that at least one of the long-term keys belonging to each of the parties involved must remain uncorrupted until the stage is completed, to avoid impersonation attacks which are otherwise trivial.

\updated{
\begin{theorem}[Post-Quantum Security of \VMuckle]
\label{thm}
    Let $\prfeval\colon \mathcal{S} \times D \to \mathsf{R}$ be a post-quantum dual PRF such that $\mathsf{R} \subseteq \mathcal{S},D$. Let $\DSIG$ be a post quantum $\EUFCMA$ secure signature scheme, $\MAC$ be a post quantum $\EUFCMA$ secure MAC with keyspace $\mathcal{K}_{\MAC} \supseteq \mathsf{R}$, $\psk \in \mathcal{S}$ be a uniformly random pre-shared key, $\KEM_{pq}$ be a post-quantum $\INDCPA$ secure KEM with keyspace $\mathcal{K}_{pq} \in \mathcal{S}$, and $k_q \in \mathcal{S}$ be a symmetric key obtained via QKD. Then the $\VMuckle$ key exchange protocol is post-quantum HAKE secure with the cleanness predicate $\clean_{\VM}$.
\end{theorem}
}

\textit{Proof.} The proof follows very closely to the security proof for $\SigMuckle$. We split the proof into cases where the query $\Test(i,s,t)$ as been issued and prove each one separately:
\begin{enumerate}
    \item The session $\pi^s_i$ (where $\pi^s_i .\rho = init$) has no origin session in stage $t$.
    \item The session $\pi^s_i$ (where $\pi^s_i.\rho = resp$) has no origin session in stage $t$.
    \item The session $\pi^s_i$ in stage $t$ has a matching session.
\end{enumerate}
\updated{Let \advA be a QPT adversary. Each proof will take the form of a series of games, where $S_n$ denotes the probability of} $\advA$ winning Game $n$. By summing the probabilities over all cases, we establish an overall bound on the advantage of $\advA$ in winning the HAKE key indistinguishability experiment. We note that the key tested in this experiment is the master secret $MS$.
    
    \subsection*{Case 1: Test $\mathsf{init}$ session without origin session.} 
    In case 1 we assume that at least one of the long term secrets (either $\psk$ or $\sk_I$) has not been corrupted, and show that an adversary $\advA$ has negligible advantage of guessing the test bit in the key indistinguishability experiment. In order for $\advA$ to win the experiment, a test key must be obtained using the $\Test$ query. As this is only possible when the session is in the $\mathsf{accept}$ state, it is sufficient to show that $\advA$ has negligible chance of getting a session to reach the $\mathsf{accept}$ state.

    \subsubsection*{Subcase 1.1: No $\CorruptSK$ query has been issued.}
    \begin{description}
        \item[Game 0:] Standard HAKE-experiment with advantage
        \begin{equation*}
            \Adv^{\mathsf{HAKE},\clean_{\VM},C_{1.1}}_{\advA,\VM,n_P,n_S,n_T}(\secpar) = \Pr[S_0].
        \end{equation*}
        \item[Game 1:] In this game, the adversary must guess the parameters $(i,s,t,j)$ corresponding to the test session $\pi_i^s$ in stage $t$ with $\pi_i^s.pid =j$. If a $\Test(i',s',t')$ query is ever received for a session with parameters $(i',s',t',j') \neq (i,s,t,j)$, the game aborts. The advantage is
        \begin{equation*}
            \Pr[S_0] \leq n^2_P n_S n_T \cdot \Pr[S_1].
        \end{equation*}
        \item[Game 2:] In Game 2, we add the rule that the game aborts if the test session $\pi_i^s$ ever reaches the status $\mathsf{reject} \gets \pi_i^s.\alpha [t]$. Since the $\Test$ query returns $\bot$ whenever this state occurs, $\advA$ cannot win. Consequently there is no difference in advantage between Game 1 and Game 2 and we have that
        \begin{equation*}
            \Pr[S_1] \leq \Pr[S_2].
        \end{equation*}
        \item[Game 3:] In Game 3, we define an event $\boldsymbol{\alpha}$ which occurs if the session reaches the status $\mathsf{accept} \gets \pi_i^s.[t]$, and causes the game to abort. Since either the game aborts before the $\mathsf{Test}$ query can be issued or the $\mathsf{Test}$ query returns $\bot$, we have $\Pr[S_3]=0$. So we only need to consider the advantage $\advA$ has in triggering the $\boldsymbol{\alpha}$ event, which is
        \begin{equation*}
            \Pr[S_{2}] \leq  \Pr[\boldsymbol{\alpha}].
        \end{equation*}  
        
        \item[Game 4:] In Game 4, we replace the computations of the $f_C \gets \prfeval(\psk, fk_C)$ and $f_S \gets \prfeval(\psk, fk_S)$ values generated so far in the session with uniformly random values $f_C' \gets \mathsf{R}$ and $f_C' \gets \mathsf{R}$ (where $\mathsf{R}$ is the output space of $\prfeval$). This is done by initializing a PRF challenger and querying the $fk_C$ and $fk_S$ values. The test values returned by the challenger are used as replacements for the $f_C$ and $f_S$ values. If the randomly generated values are returned, then we are in Game 4, otherwise we are in Game 3. By our initial assumption no $\CorruptSK$ queries have been issued, so this is a valid substitution. Any adversary that can differentiate between these two games can be turned into a successful PRF distinguisher. Therefore, the advantage is
        \begin{equation*}
            \Pr[\boldsymbol{\alpha}] \leq \Pr[S_4] + \Adv^{\mathsf{prf}}_{\advA,\prfeval}(\secpar).
        \end{equation*}

        \item[Game 5:] In Game 5, we define another abort event which is triggered if the session $\pi_i^s.[t]$ receives a $\MAC$ authentication code $\tau_I$ that verifies correctly. We do this by initializing an $\EUFCMA$ $\MAC$ challenger, and replacing $f_C'$ with the $\sk$ value generated by the challenger. Since $f_C'$ is already uniformly random by Game 4 this is a valid substitution. We then submit any $\MAC$ code $\tau_I$ received during the session to the challenger as a forgery on the message $H_6$. If $\tau_I$ is verified, then the abort event is triggered and $\advA$ does not win. If no authentication code is received, or if $\tau_I$ is not verified correctly, the session will not reach the $\mathsf{accept}$ stage, so $\Pr[S_5]=0$. Thus the adversary can only win Game 4 if they are capable of producing a valid authentication tag on the message $H_6$ without knowing the secret key $\sk$. Any adversary capable of doing this is a successful $\EUFCMA$ adversary against $\MAC$. The resulting bound is
        \begin{equation*}
            \Pr[S_4] \leq  \AdvEUFCMA{\advA, \MAC}(\secpar).
        \end{equation*}
        
    \end{description}
    Hence, we obtain the following advantage
    \begin{equation*}
        \Pr[S_0] \leq n^2_P n_S n_T \cdot \left(
            \Adv^{\mathsf{prf}}_{\advA,\prfeval}(\secpar) 
            + \AdvEUFCMA{\advA,\MAC}(\secpar) 
        \right).
    \end{equation*}
        
    \subsubsection*{Subcase 1.2: No $\CorruptQK$ has been issued.} 
    In this case, we assume that the signing key $\sk_I$ is not corrupted, and show that the adversary has negligible advantage in getting the test session to reach the $\mathsf{accept}$ state. The proof follows along the lines of the proof of case 1 in \cite{BruRamStr23}, and we do not repeat it here. The advantage is
    \begin{equation*}
        \Pr[S_0] \leq n^2_P n_S n_T \cdot \left(\AdvEUFCMA{\advA,\DSIG}(\secpar)\right).
    \end{equation*}
        
    \paragraph*{} Summing the advantages of the two subcases, we establish a bound on the overall advantage of $\advA$ in Case 1:
    \begingroup
        \fontsize{8.6pt}{12pt}\selectfont
        \begin{equation*}
            \Pr[S_0] \leq n^2_P n_S n_T \cdot \left(
                \Adv^{\mathsf{prf}}_{\advA,\prfeval}(\secpar)
                + \AdvEUFCMA{\advA,\MAC}(\secpar) 
                + \AdvEUFCMA{\advA,\DSIG}(\secpar) 
            \right).
        \end{equation*}
    \endgroup
        
    \subsection*{Case 2: Test $\mathsf{resp}$ session without origin session.} 
    This case considers whether the adversary is able to win the key indistinguishability experiment given the responder role. The proof follows analogously to the proof for Case 1 and results in the same advantage of
    \begingroup
        \fontsize{8.6pt}{12pt}\selectfont
        \begin{equation*}
            \Pr[S_0] \leq n^2_P n_S n_T \cdot \left(
                \Adv^{\mathsf{prf}}_{\advA,\prfeval}(\secpar)
                + \AdvEUFCMA{\advA,\MAC}(\secpar) 
                + \AdvEUFCMA{\advA,\DSIG}(\secpar) 
            \right).
        \end{equation*}
    \endgroup

    \subsection*{Case 3: Test session with matching session.} 
    Here we show that an adversary $\advA$ has negligible advantage in winning the key-indistinguishability experiment for a session which has a matching session provided that at least one of the ephemeral session keys has not been revealed. We look at 4 cases covering each possibility, and prove each through a series of games. 

    \subsubsection*{Subcase 3.1: No $\CompromiseQK(i,s,t)$ or $\CompromiseQK(j,r,t)$ have been issued.}
    This case shows that an adversary issuing a $\Test$ query has negligible advantage in winning the key-indistinguishability game if the post-quantum key is not compromised. We may assume that all other ephemeral and long term secrets are known to the adversary. 
    The proof for this follows the proof of case 3.1 in \cite{BruRamStr23}, with the advantage
    \begingroup
        \fontsize{8.2pt}{12pt}\selectfont
        \[
        \Pr[S_0] \leq n^2_P n^2_S n_T \cdot \left( 
            5 \cdot \Adv^{\mathsf{prf}}_{\advA,\prfeval}(\secpar) 
            + 3 \cdot \Adv^{\mathsf{dual\text{-}prf}}_{\advA,\prfeval}(\secpar)
            + \Adv^{\mathsf{ind\text{-}cpa}}_{\advA,\KEM}(\secpar)             
        \right).
        \]
    \endgroup

    \subsubsection*{Subcase 3.2: No $\CompromiseSK(i,s,t)$ or $\CompromiseSK(j,r,t)$ have been issued.} 
    This case shows, that if the attacker issues a $\Test$ query to a session that is clean due to the secrecy of the ephemeral quantum key, the attacker has a negligible advantage in guessing the test bit. In this scenario, all ephemeral secrets except the quantum key as well as the long-term classical and post-quantum secrets are known to the attacker.
    The proof for this case follows along the lines of of case 3.2 in \cite{BruRamStr23}, and the advantage is
    \[
    \Pr[S_0] \leq n^2_P n^2_S n_T \cdot \left( 
        4 \cdot \Adv^{\mathsf{prf}}_{\advA,\prfeval}(\secpar) 
        + \Adv^{\mathsf{dual\text{-}prf}}_{\advA,\prfeval}(\secpar)
    \right).
    \]

    \subsubsection*{Subcase 3.3: No $\CompromiseQK(i, s, t')$ or $\CompromiseQK(j, r, t')$ have been issued with $\pi^s_i$ matching $\pi^r_j$ in stages $u$ where $t' \leq u \leq t$. No $\CompromiseSS(i, s, u)$, $\CompromiseSS(j, r, u)$ have been issued (except possibly in stage $t'$). No $\Reveal(i,s,u)$ or $\Reveal(j,r,u)$ have been issued.}
    This case shows, that if a previous session has been completed cleanly under the predicate $\clean_{\VM}$ and $\advA$ has not compromised the session state \SecState since then, the attacker has a negligible advantage in guessing the test bit of the current session.
    The proof for this case follows the proof of case 3.3 in \cite{BruRamStr23} 
    \footnote{In this paper we use a different definition of the cleanness predicate than that used in the Muckle+ paper. This results in a slightly tighter bound on the $\mathsf{prf}$ advantages in subcases 3.3 and 3.4.}. The resultant advantage is
    \begingroup
        \fontsize{7.2pt}{12pt}\selectfont
        \[
        \Pr[S_0] \leq n^2_P n^2_S n^2_T \cdot \left( 
            (1+4 n_T) \cdot \Adv^{\mathsf{prf}}_{\advA,\prfeval}(\secpar) 
            + 3 \cdot \Adv^{\mathsf{dual\text{-}prf}}_{\advA,\prfeval}(\secpar)
            + \Adv^{\mathsf{ind\text{-}cpa}}_{\advA,\KEM}(\secpar) 
        \right).
        \]
    \endgroup

    \subsubsection*{Subcase 3.4: No $\CompromiseSK(i, s, t')$ or $\CompromiseSK(j, r, t')$ have been issued with $\pi^s_i$ matching $\pi^r_j$ in stages $u$ where $t' \leq u \leq t$. No $\CompromiseSS(i, s, u)$ or $\CompromiseSS(j, r, u)$ have been issued (except possibly in stage $t'$). No $\Reveal(i,s,u)$ or $\Reveal(j,r,u)$ have been issued.}
    In this case we suppose that the quantum key generated in a previous session is not compromised, and as a result that that session is clean under the predicate $\clean_{\VM}$, and that the $\SecState$ for all sessions since then was not compromised. Under this assumption, we show that $\advA$ has a negligible advantage in determining the current $\SecState$, and consequently of determining the test bit.
    The proof for this case is similar to the proof of case 3.4 in \cite{BruRamStr23}, with the advantage
    \[
    \Pr[S_0] \leq n^2_P n^2_S n^2_T \cdot \left( 4 n_T \cdot \Adv^{\mathsf{prf}}_{\advA,\prfeval}(\secpar) + \Adv^{\mathsf{dual\text{-}prf}}_{\advA,\prfeval}(\secpar)\right).
    \]

    \subsection*{Final Derivation}
    By taking the sum of the advantages in each subcase, we can determine a bound on the overall advantage an adversary has in winning the HAKE security experiment under the conditions specified in the theorem. The result is the following:
    \begingroup
        \fontsize{8.3pt}{12pt}\selectfont
        \begin{equation*}
            \begin{array}{l}
                \Adv^{\mathsf{HAKE},\clean_{\VM}}_{\advA,\VM,n_P,n_S,n_T}(\secpar) \leq \\
                2 \cdot n^2_P n_S n_T \cdot \left(
                    \Adv^{\mathsf{prf}}_{\advA,\prfeval}(\secpar)
                    + \AdvEUFCMA{\advA,\MAC}(\secpar) 
                    + \AdvEUFCMA{\advA,\DSIG}(\secpar)
                \right) + \\
                n^2_P n^2_S n_T \cdot \left( 
                    9 \cdot \Adv^{\mathsf{prf}}_{\advA,\prfeval}(\secpar)
                    + 4 \cdot \Adv^{\mathsf{dual\text{-}prf}}_{\advA,\prfeval}(\secpar)
                    + \Adv^{\mathsf{ind\text{-}cpa}}_{\advA,\KEM}(\secpar) 
                \right) + \\
                n^2_P n^2_S n^2_T \cdot \left( 
                    (1+8 n_T) \cdot \Adv^{\mathsf{prf}}_{\advA,\prfeval}(\secpar) 
                    + 4 \cdot \Adv^{\mathsf{dual\text{-}prf}}_{\advA,\prfeval}(\secpar)
                    + \Adv^{\mathsf{ind\text{-}cpa}}_{\advA,\KEM}(\secpar) 
                \right).
            \end{array}
        \end{equation*}
    \endgroup

    Since each component of this bound is negligible according to our initial assumptions, we conclude that a \updated{QPT} adversary \updated{has negligible advantage} in winning the \updated{post-quantum} HAKE indistinguishability experiment.

\updated{
\begin{corr}
\label{corr1}
    Let $\prfeval\colon \mathcal{S} \times D \to \mathsf{R}$, $\DSIG$, $\MAC$, $\psk$, and $k_q$ be defined as in Theorem \ref{thm}, replacing the post-quantum security assumptions with their classical equivalents. In addition, let $\KEM_c$ be an $\INDCPA$ secure $\KEM$ with keyspace $\mathcal{K}_c \subseteq \mathcal{S}$. Then the \SigMuckle key exchange protocol is HAKE secure with the cleanness predicate $\clean_{c\VM}$.
\end{corr}
}

\textit{Proof.} The proof can easily be adapted from the proof of Theorem \ref{thm}.

\updated{
\begin{remark}
    In Theorem \ref{thm} we assumed the security of all components, and proved HAKE security according to the cleanness predicate $\clean_\VM$. To consider the security in case some components do not meet these assumptions, we simply assume that all associated keys are revealed and restrict the cleanness predicate to exclude the cases that rely on the security of these keys. Similarly, we can adapt the classical result from Cor. \ref{corr1} by modifying the predicate $\clean_{c\VM}$.
\end{remark}
}


\begin{thebibliography}{10}
\providecommand{\url}[1]{#1}
\csname url@samestyle\endcsname
\providecommand{\newblock}{\relax}
\providecommand{\bibinfo}[2]{#2}
\providecommand{\BIBentrySTDinterwordspacing}{\spaceskip=0pt\relax}
\providecommand{\BIBentryALTinterwordstretchfactor}{4}
\providecommand{\BIBentryALTinterwordspacing}{\spaceskip=\fontdimen2\font plus
\BIBentryALTinterwordstretchfactor\fontdimen3\font minus \fontdimen4\font\relax}
\providecommand{\BIBforeignlanguage}[2]{{%
\expandafter\ifx\csname l@#1\endcsname\relax
\typeout{** WARNING: IEEEtran.bst: No hyphenation pattern has been}%
\typeout{** loaded for the language `#1'. Using the pattern for}%
\typeout{** the default language instead.}%
\else
\language=\csname l@#1\endcsname
\fi
#2}}
\providecommand{\BIBdecl}{\relax}
\BIBdecl

\bibitem{Shor}
\BIBentryALTinterwordspacing
P.~W. Shor, ``{Polynomial-Time Algorithms for Prime Factorization and Discrete Logarithms on a Quantum Computer},'' \emph{SIAM J. Comput.}, vol.~26, no.~5, p. 1484–1509, October 1997. [Online]. Available: \url{https://doi.org/10.1137/S0097539795293172}
\BIBentrySTDinterwordspacing

\bibitem{NIST_8413}
G.~Alagic, D.~Cooper, Q.~Dang, T.~Dang, J.~M. Kelsey, J.~Lichtinger, Y.-K. Liu, C.~A. Miller, D.~Moody, R.~Peralta, R.~Perlner, A.~Robinson, D.~Smith-Tone, and D.~Apon, ``\BIBforeignlanguage{en}{{Status Report on the Third Round of the NIST Post-Quantum Cryptography Standardization Process}},'' NIST IR 8413-upd1, Tech. Rep., July 2022.

\bibitem{FIPS203}
N.~I. of~Standards and Technology, ``{Module-Lattice-Based Key-Encapsulation Mechanism Standard},'' U.S. Department of Commerce, Washington, D.C., Tech. Rep. Federal Information Processing Standards Publications (FIPS PUBS) 203, August 2024.

\bibitem{FIPS204}
------, ``{Module-Lattice-Based Digital Signature Standard},'' U.S. Department of Commerce, Washington, D.C., Tech. Rep. Federal Information Processing Standards Publications (FIPS PUBS) 204, August 2024.

\bibitem{FIPS205}
------, ``{Stateless Hash-Based Digital Signature Standard},'' U.S. Department of Commerce, Washington, D.C., Tech. Rep. Federal Information Processing Standards Publications (FIPS PUBS) 205, August 2024.

\bibitem{QKD}
\BIBentryALTinterwordspacing
N.~Gisin, G.~Ribordy, W.~Tittel, and H.~Zbinden, ``Quantum cryptography,'' \emph{Rev. Mod. Phys.}, vol.~74, pp. 145--195, Mar 2002. [Online]. Available: \url{https://link.aps.org/doi/10.1103/RevModPhys.74.145}
\BIBentrySTDinterwordspacing

\bibitem{PQC}
\BIBentryALTinterwordspacing
D.~J. Bernstein and T.~Lange, ``Post-quantum cryptography---dealing with the fallout of physics success,'' Cryptology {ePrint} Archive, Paper 2017/314, 2017. [Online]. Available: \url{https://eprint.iacr.org/2017/314}
\BIBentrySTDinterwordspacing

\bibitem{Scarani}
\BIBentryALTinterwordspacing
V.~Scarani, H.~Bechmann-Pasquinucci, N.~J. Cerf, M.~Du\v{s}ek, N.~L\"utkenhaus, and M.~Peev, ``The security of practical quantum key distribution,'' \emph{Rev. Mod. Phys.}, vol.~81, pp. 1301--1350, Sep 2009. [Online]. Available: \url{https://link.aps.org/doi/10.1103/RevModPhys.81.1301}
\BIBentrySTDinterwordspacing

\bibitem{QS_MACsec}
S.-L. Gazdag, S.~Grundner-Culemann, T.~Heider, D.~Herzinger, F.~Sch{\"a}rtl, J.~Y. Cho, T.~Guggemos, and D.~Loebenberger, ``{Quantum-Resistant MACsec and IPsec for Virtual Private Networks},'' in \emph{Security Standardisation Research}, F.~G{\"u}nther and J.~Hesse, Eds.\hskip 1em plus 0.5em minus 0.4em\relax Cham: Springer Nature Switzerland, 2023, pp. 1--21.

\bibitem{TLS_ArXiV}
J.~S. Buruaga, R.~B. Méndez, J.~P. Brito, and V.~Martin, ``Quantum-safe integration of tls in sdn networks,'' 2025.

\bibitem{YAF24}
Z.~Yang, H.~Alfauri, B.~Farkiani, R.~Jain, R.~D. Pietro, and A.~Erbad, ``A survey and comparison of post-quantum and quantum blockchains,'' \emph{IEEE Communications Surveys \& Tutorials}, vol.~26, no.~2, pp. 967--1002, 2024.

\bibitem{YSJ22}
Z.~Yang, T.~Salman, R.~Jain, and R.~D. Pietro, ``Decentralization using quantum blockchain: A theoretical analysis,'' \emph{IEEE Transactions on Quantum Engineering}, vol.~3, pp. 1--16, 2022.

\bibitem{IEEE802.1AE-2018}
``{IEEE Standard for Local and Metropolitan Area Networks-Media Access Control (MAC) Security},'' \emph{{IEEE Std 802.1AE-2018 (Revision of IEEE Std 802.1AE-2006)}}, pp. 1--239, 2018.

\bibitem{OSI}
``{Information technology — Open Systems Interconnection — Basic Reference Model: The Basic Model},'' International Organization for Standardization, Geneva, CH, Standard, 11 1994.

\bibitem{AES_QR}
\BIBentryALTinterwordspacing
X.~Bonnetain, M.~Naya-Plasencia, and A.~Schrottenloher, ``{Quantum Security Analysis of AES},'' \emph{{IACR Transactions on Symmetric Cryptology}}, vol. 2019, no.~2, pp. 55--93, Jun. 2019. [Online]. Available: \url{https://inria.hal.science/hal-02397049}
\BIBentrySTDinterwordspacing

\bibitem{IEEE802.1X}
``{IEEE Standard for Local and Metropolitan Area Networks-Port-Based Network Access Control},'' \emph{IEEE Std 802.1X-2020}, 2020.

\bibitem{Stebila_24}
\BIBentryALTinterwordspacing
D.~Stebila, S.~Fluhrer, and S.~Gueron, ``{Hybrid key exchange in TLS 1.3},'' Internet Engineering Task Force, Internet-Draft draft-ietf-tls-hybrid-design-11, Oct. 2024, work in Progress. [Online]. Available: \url{https://datatracker.ietf.org/doc/draft-ietf-tls-hybrid-design/11/}
\BIBentrySTDinterwordspacing

\bibitem{BruRamStr23}
S.~Bruckner, S.~Ramacher, and C.~Striecks, ``{Muckle+: End-to-End Hybrid Authenticated Key Exchanges},'' in \emph{Post-Quantum Cryptography}, T.~Johansson and D.~Smith-Tone, Eds.\hskip 1em plus 0.5em minus 0.4em\relax Cham: Springer Nature Switzerland, 2023, pp. 601--633.

\bibitem{DowHanPat20}
B.~Dowling, T.~B. Hansen, and K.~G. Paterson, ``{Many a Mickle Makes a Muckle: A Framework for Provably Quantum-Secure Hybrid Key Exchange},'' in \emph{Post-Quantum Cryptography}, J.~Ding and J.-P. Tillich, Eds.\hskip 1em plus 0.5em minus 0.4em\relax Cham: Springer International Publishing, 2020, pp. 483--502.

\bibitem{EuroQCI}
E.~Comission, ``{The European Quantum Communication Infrastructure (EuroQCI) Initiative},'' 10 2024, \url{https://digital-strategy.ec.europa.eu/en/policies/european-quantum-communication-infrastructure-euroqci} [Accessed: (29/10/2024)].

\bibitem{PQC_MACsec}
\BIBentryALTinterwordspacing
J.~Y. Cho and A.~Sergeev, ``{Post-quantum MACsec in Ethernet Networks},'' \emph{Journal of Cyber Security and Mobility}, vol.~10, no.~1, p. 161–176, Mar. 2021. [Online]. Available: \url{https://journals.riverpublishers.com/index.php/JCSANDM/article/view/5973}
\BIBentrySTDinterwordspacing

\bibitem{QKD_MACsec}
\BIBentryALTinterwordspacing
------, ``{Using QKD in MACsec for secure Ethernet networks},'' \emph{IET Quantum Communication}, vol.~2, no.~3, pp. 66--73, 2021. [Online]. Available: \url{https://ietresearch.onlinelibrary.wiley.com/doi/abs/10.1049/qtc2.12006}
\BIBentrySTDinterwordspacing

\bibitem{JC:DFGS21}
B.~Dowling, M.~Fischlin, F.~G{\"u}nther, and D.~Stebila, ``A cryptographic analysis of the {TLS} 1.3 handshake protocol,'' \emph{Journal of Cryptology}, vol.~34, no.~4, p.~37, Oct. 2021.

\bibitem{SteMos16}
D.~Stebila and M.~Mosca, ``{Post-quantum Key Exchange for the Internet and the Open Quantum Safe Project},'' in \emph{Selected Areas in Cryptography -- SAC 2016}, R.~Avanzi and H.~Heys, Eds.\hskip 1em plus 0.5em minus 0.4em\relax Cham: Springer International Publishing, 2017, pp. 14--37.

\bibitem{battarbee2024quantumsafehybridkeyexchanges}
\BIBentryALTinterwordspacing
C.~Battarbee, C.~Striecks, L.~Perret, S.~Ramacher, and K.~Verhaeghe, ``Quantum-safe hybrid key exchanges with kem-based authentication,'' 2024. [Online]. Available: \url{https://arxiv.org/abs/2411.04030}
\BIBentrySTDinterwordspacing

\bibitem{BDLSY11}
D.~J. Bernstein, N.~Duif, T.~Lange, P.~Schwabe, and B.-Y. Yang, ``{High-Speed High-Security Signatures},'' in \emph{Cryptographic Hardware and Embedded Systems -- CHES 2011}, B.~Preneel and T.~Takagi, Eds.\hskip 1em plus 0.5em minus 0.4em\relax Berlin, Heidelberg: Springer Berlin Heidelberg, 2011, pp. 124--142.

\bibitem{RFC_8446}
E.~Rescorla, ``{The Transport Layer Security (TLS) Protocol Version 1.3},'' RFC 8446, ago 2018.

\bibitem{RFC_9190}
J.~P. Mattsson and M.~Sethi, ``{EAP-TLS 1.3: Using the Extensible Authentication Protocol with TLS 1.3},'' RFC 9190, feb 2022.

\bibitem{MadQCI}
\BIBentryALTinterwordspacing
V.~Martin, J.~P. Brito, L.~Ortiz, R.~B. Mendez, J.~S. Buruaga, R.~J. Vicente, A.~Sebastián-Lombraña, D.~Rincon, F.~Perez, C.~Sanchez, M.~Peev, H.~H. Brunner, F.~Fung, A.~Poppe, F.~Fröwis, A.~J. Shields, R.~I. Woodward, H.~Griesser, S.~Roehrich, F.~D.~L. Iglesia, C.~Abellan, M.~Hentschel, J.~M. Rivas-Moscoso, A.~Pastor, J.~Folgueira, and D.~R. Lopez, ``{MadQCI: a heterogeneous and scalable SDN-QKD network deployed in production facilities},'' \emph{npj Quantum Information}, vol.~10, no.~1, p.~80, Sep 2024. [Online]. Available: \url{https://doi.org/10.1038/s41534-024-00873-2}
\BIBentrySTDinterwordspacing

\bibitem{EPRINT:BelLys15}
M.~Bellare and A.~Lysyanskaya, ``Symmetric and dual {PRFs} from standard assumptions: {A} generic validation of an {HMAC} assumption,'' Cryptology ePrint Archive, Report 2015/1198, 2015, \url{https://eprint.iacr.org/2015/1198}.

\bibitem{RBBK01}
P.~Rogaway, M.~Bellare, J.~Black, and T.~Krovetz, ``{OCB: a block-cipher mode of operation for efficient authenticated encryption},'' in \emph{Proceedings of the 8th ACM Conference on Computer and Communications Security}, ser. CCS '01.\hskip 1em plus 0.5em minus 0.4em\relax New York, NY, USA: Association for Computing Machinery, 2001, p. 196–205.

\bibitem{LKZC07}
J.~Li, K.~Kim, F.~Zhang, and X.~Chen, ``Aggregate proxy signature and verifiably encrypted proxy signature,'' in \emph{Provable Security}, W.~Susilo, J.~K. Liu, and Y.~Mu, Eds.\hskip 1em plus 0.5em minus 0.4em\relax Berlin, Heidelberg: Springer Berlin Heidelberg, 2007, pp. 208--217.

\bibitem{CreFel12}
C.~Cremers and M.~Feltz, ``{Beyond eCK: Perfect Forward Secrecy under Actor Compromise and Ephemeral-Key Reveal},'' in \emph{Computer Security -- ESORICS 2012}, S.~Foresti, M.~Yung, and F.~Martinelli, Eds.\hskip 1em plus 0.5em minus 0.4em\relax Berlin, Heidelberg: Springer Berlin Heidelberg, 2012, pp. 734--751.

\end{thebibliography}

\end{document}